\documentclass[11pt]{article}

\usepackage{color,graphicx}
\usepackage{young}
\usepackage{caption}
\usepackage[vcentermath]{youngtab}
\usepackage{amsmath,amssymb,graphicx}
\usepackage{hyperref}
\definecolor{darkred}{rgb}{0.65,0.15,0}
\hypersetup{pdfborder={0 0 0},colorlinks=true,urlcolor=darkred,citecolor=blue,linkcolor=darkred,linktocpage=true}

\usepackage{cite}
\usepackage{multirow}
\usepackage{amsmath}
\usepackage{amsfonts}
\usepackage{amssymb}
\usepackage{graphicx}%
\usepackage{amsthm}
\usepackage{mathrsfs}
\usepackage[T1]{fontenc}
\usepackage{enumerate}
\usepackage{color}

\def\4diml{four-dimensional}

\def\-1{^{-1}}

\makeatletter

\@addtoreset{equation}{section}
\makeatother

\begin{document}

\thispagestyle{empty}

\vspace{5mm}

\begin{center}
{\LARGE \bf T-dualization of G\"{o}del string cosmologies via\\[2mm] Poisson-Lie T-duality approach}

\vspace{15mm}
\normalsize
{\large   Ali Eghbali\footnote{Corresponding author: eghbali978@gmail.com}, Reza Naderi\footnote{r\_naderi8@yahoo.com},
Adel  Rezaei-Aghdam\footnote{rezaei-a@azaruniv.ac.ir}}

\vspace{2mm}
{\small \em Department of Physics, Faculty of Basic Sciences,\\
Azarbaijan Shahid Madani University, 53714-161, Tabriz, Iran}\\
\vspace{4mm}


\vspace{6mm}

\begin{tabular}{p{12cm}}
{\small
Using the homogeneous G\"{o}del spacetimes we find some new solutions for the field equations of
bosonic string effective action up to  first order in $\alpha'$ including both dilaton and axion fields.
We then discuss in detail the (non-)Abelian T-dualization of G\"{o}del string cosmologies via the Poisson-Lie (PL) T-duality approach.
In studying Abelian T-duality of the models we get seven dual models
in such a way that they are constructed by one-, two- and three-dimensional
Abelian Lie groups acting freely on the target space manifold.
The results of our study show that the Abelian T-dual models are, under some of the special conditions, self-dual;
moreover, by applying the usual rules of Abelian
T-duality without further corrections, we are still able to obtain two-loop solutions.
We also study the Abelian T-duality of G\"{o}del string cosmologies up to $\alpha'$-corrections by using
the T-duality rules at two-loop order derived by Kaloper and Meissner.
Afterwards, non-Abelian duals of the G\"{o}del spacetimes are constructed by two- and three-dimensional non-Abelian Lie groups
such as $A_2$, $A_2 \oplus A_1$ and $SL(2, \mathbb{R})$.
In this way, the PL self-duality of $AdS_3 \times \mathbb{R}$ space is discussed.
}
\end{tabular}
\vspace{-1mm}
\end{center}


\setcounter{page}{1}

\newpage

\tableofcontents

 \vspace{5mm}

\section{Introduction}
\label{sec:intro}
One of the most interesting cosmological solutions to Einstein's field
equations is G\"{o}del  spacetime \cite{godel} which constituted (and still
constitutes) a considerable motivation to the investigation
of solutions more complex than those treated until then.
It was obtained \cite{Barrow} some new solutions of string theory, including terms up to the first
order in the inverse string tension $\alpha'$ for the homogeneous G\"{o}del spacetimes.
Recently, it has been shown that \cite{H.Lu} four-dimensional G\"{o}del universe can be embedded in string
theory. The corresponding  Lagrangians  to the
Einstein-Maxwell-Axion, Einstein-Proca-Axion  and Freedman-Schwarz  $SU(2) \times
SU(2)$ gauged supergravity theories admit the G\"{o}del metrics as
solutions, all involving only the fundamental matter fields \cite{H.Lu} (see also \cite{Israeil}).
In Ref. \cite{Barrow}, to find a class of G\"{o}del universes without  Closed
Timelike Curves (CTC's) within the framework of low-energy effective
string theory, it has been considered a convenient ansatz for both dilaton and
axion fields in an orthonormal frame. Thus, the G\"{o}del spacetimes
are of particular importance both in general relativity and string theory.
But so far their target space duality (T-duality) have not been studied.
In the present work,
we obtain the other forms of solutions of equations (but not in an orthonormal frame) for
the two-loop beta-function including the G\"{o}del spacetimes, field strength and dilaton field.
The existence of solution for beta-function equations possessing
the G\"{o}del metrics with appropriate axion and dilaton fields, helps us to study their T-dual spaces.
We furthermore get new solutions by considering a simpler form of G\"{o}del metrics,
and then focus on finding target space duals of the solutions. Accordingly, we improve Barrow's results  \cite{Barrow}.

T-duality is a very important symmetry of string theory which was originally
defined for a string theory $\sigma$-model where the backgrounds of
model have an Abelian group of isometries \cite{{Buscher1},{Buscher2}}.
T-duality is a peculiar feature of strings, since it provides a method for relating seemingly inequivalent string theories, and
allows to build new string backgrounds which could not be addressed otherwise.
The theory of Abelian T-duality is well understood and had been the subject of much research (for a review see
e.g. \cite{porrati}).
Then, the basic duality procedure could be generalized to the
case where the original $\sigma$-model had a non-Abelian group of
isometries \cite{Ossa} (further work in this direction was carried out
in \cite{{Rocek},{Gasperini},{Givoen},{Alvarez},{Tyurin1},{Elitzur}}).
Whereas in the Abelian case, the dual theory has a natural symmetry suitable for inverting
the duality transformation, in the non-Abelian case it has been shown that \cite{Givoen} this symmetry
becomes nonlocal and cannot obviously be used for the inverse. In fact, it has been found
that the non-Abelian T-duality is not an exact symmetry of the conformal field theory, and conjectured
that it is a map between different conformal field theories.
There has been a new interest in  non-Abelian T-duality, which was ignited by \cite{Sfetsos0},
that provided the transformation rule for the Ramond-Ramond fields under the non-Abelian T-duality.
It was then extended to $\sigma$-models with nonvanishing Ramond fluxes,
thus allowing to search for new supergravity solutions \cite{{Lozano},{Nunez1},{Sfet.},{Nunez2},{Nunez3}}.
Klim\v{c}ik and \v{S}evera
proposed a generalization of T-duality, or the so-called  PL T-duality \cite{{Klim1},{Klim2}},
which allows the duality to be performed on a target space without isometries.
Afterwards, PL T-duality transformations could be generalized to the Lie supergroups \cite{ER1}, as well as supermanifolds \cite{ER5},
in such a way that super PL symmetry of the WZW models based on some of the Lie supergroups of superdimension $(2|2)$
was studied \cite{{ER7},{ER8}}. Of course, in the Lie groups case, PL symmetry of the
WZW models was already been studied in Ref. \cite{Alekseev1} (see also \cite{{Sfetsos1},{Lledo},{eghbali11}}).
Recently, PL T-duality also appears as an important tool in
the study of integrable models and their deformations \cite{{Wulff1},{Wulff2},{Hoare1},{Klim3},{Vicedo},{Hoare2},{Sfetsos3}}.
Lately, using the PL T-duality approach in the presence of spectator fields we have found new dual solutions for some of the gravitational and string
backgrounds such as BTZ black hole \cite{EMR13} and the WZW models based on the Lie groups $H_4$
and $GL(2,\mathbb{R})$ \cite{Eghbali.2}.

The main purpose of this paper is to study the (non-)Abelian T-dualization (here as the PL T-duality on a (semi-)Abelian double)  of
the G\"{o}del string cosmologies. The procedure for this study is as follows:
in the context of PL T-duality with spectators, the choice of spectator-dependent matrices
$(E_{0}, F, F^{{(1)}}, F^{{(2)}})$  plays a key role in the process.
Our choice is usually  based on the solutions of the vanishing of beta-function equations.
By a suitable choice, we obtain the background of the original $\sigma$-model including the G\"{o}del metrics and
B-field (corresponding to a constant torsion).
We then use the PL T-duality transformations in order to get the corresponding dual solutions.
Also, the duality transformation of the dilaton field is, at one-loop level, given by equations \eqref{c.16.1} and \eqref{c.16.2}.
Finally, we are interested in testing the conformal invariance of the dual solutions up to the two-loop order.
{\it We notice that in the case of the G\"{o}del spacetimes one can apply the usual rules of (non-)Abelian
T-duality without further corrections, and still be able to obtain two-loop
solutions. In general, further corrections to the rules are
necessary.} Recently, it has been shown that \cite{Wulff22} (see also \cite{Hassler2}) the PL duality can be extended to order
$\alpha'$, i.e. two loops in the $\sigma$-model perturbation theory,
provided that the map is corrected.
Furthermore, by using the higher derivative formulation of DFT,
it has been found \cite{Marques} a unified expression for first order corrections to generalized
dualities so that it can be easily specified to any generalized T-duality (Abelian, non-Abelian,
PL, etc.).

As explained above, we obtain  the Abelian T-duals of
the G\"{o}del metrics by using the PL T-duality approach and then testing the conformal invariance conditions of the duals
up to two-loop order. When going to two-loop it
is known that in general the rules of Abelian T-duality must
receive corrections.
The two-loop $\sigma$-model corrections to the Abelian T-duality map were obtained by Kaloper and Meissner (KM) in \cite{KM}.
They had used the effective action approach by focusing on backgrounds that
have a single Abelian isometry.
By using the T-duality rules of KM,
we study the Abelian T-duality of G\"{o}del string cosmologies up to $\alpha'$-corrections
when the dualizing is implemented by the shift of directions $z$ and $t$.


This paper is organized as follows. After the introduction section,  Sec. \ref{Sec.II} reviews the
conformal invariance conditions  of the $\sigma$-model up to  the first
order in  $\alpha'$.
We start Sec. \ref{Sec.III} by introducing the G\"{o}del metrics and
then discuss the solutions of two-loop beta-function equations
possessing the G\"{o}del spacetimes.
A short review of PL T-dual $\sigma$-models construction in the presence of spectator fields is presented in Sec. \ref{Sec.IV},
where necessary formulas are summarized.
In Sec. \ref{Sec.V}, we study the Abelian T-dualization of the G\"{o}del spacetimes via the
PL T-duality approach.
The study of the Abelian T-duality of the G\"{o}del spacetimes up to $\alpha'$-corrections
using the KM approach, when the duality is implemented by a shift of the coordinates $z$ and $t$,
is discussed in Sec. \ref{Sec.VI}.
The non-Abelian duals of the G\"{o}del spacetimes constructed by two- and three-dimensional non-Abelian Lie groups
are given in Sec. \ref{Sec.VII}. In this section, we also study
the PL self-duality of the $AdS_3 \times \mathbb{R}$ space. The results of Abelian and non-Abelian T-duality of the 
G\"{o}del string cosmologies are clearly summarized in Tables 1, 2 and 3.
Some concluding remarks are given in the last section.

\section{Two-loop conformal invariance conditions of the bosonic string $\sigma$-model}
\label{Sec.II}

A bosonic string propagating on a non-trivial background can be described
by the well-known $\sigma$-model defined on a two-dimensional curved
surface ${\Sigma}$ in $d$ spacetime dimensions with metric ${G}_{_{MN}}$,
antisymmetric tensor field ${B}_{_{MN}}$ (axion field) and dilaton field $\Phi$
\begin{eqnarray}\label{a.1}
S &=& \frac{1}{4 \pi \alpha'}\int_{{\Sigma}}\!d\tau  d\sigma \sqrt{-h} \big[h^{\alpha \beta}G_{_{MN}}(X)
+\epsilon^{\alpha \beta} B_{_{MN}} (X)\big]\partial_{\alpha}X{^{M}}
\partial_{_\beta}X^{{N}} \nonumber\\
&&~~~~~~~~~~~~~~~~~~~~~~~~~~~~~~~~~~~~~~~~~~~~~~~~~~~~+ \frac{1}{8 \pi} \int_{{\Sigma}}\!d\tau  d\sigma  ~R^{^{(2)}} ~\Phi(X),
\end{eqnarray}
where $\sigma^{\alpha}=(\tau , \sigma)$ are the string worldsheet coordinates, and
$X^{^{M}}$ (${M} =1,...,d$) are coordinates in spacetime.
$h_{\alpha \beta}$ and $R^{^{(2)}}$ are the induced metric and curvature scalar on the string worldsheet, respectively.
$\epsilon^{\alpha \beta}$ is an antisymmetric tensor on the worldsheet and $h=\det h_{\alpha \beta}$.
The dimensionful coupling constant $\alpha'$ turns out to be the inverse string tension.

Since we are considering bosonic string theory, there is only one more massless degree
of freedom of the string, namely the dilaton $\Phi$. This gives a contribution to the action in the form of the second term of \eqref{a.1}.
This term breaks Weyl invariance on a classical level as do the one-loop corrections to ${G}$ and  ${B}$.

In the $\sigma$-model context, the conformal invariance conditions of the $\sigma$-model \eqref{a.1} are
provided by the vanishing of the beta-function equations \cite{callan}.
In order for the fields $(G, B, \Phi)$ to provide a consistent string background at low-energy
up to two-loop order (first order in $\alpha'$)
they must satisfy the following equations \cite{{A.Sen1},{A.Sen2},{Tseytlin},{c.hull},{Metsaev}}
\begin{eqnarray}
&&{\cal R}_{_{MN}}-H^2_{_{MN}}+{\nabla}_{_M}
{\nabla}_{_N} \Phi +\frac{1}{2} \alpha' \Big[{\cal R}_{_{M PQR}} {\cal R}_{_N}^{^{~PQR}}
+2 {\cal R}_{_{M PQ N}} {H^2}^{^{PQ}}\nonumber\\
&&~~~~~~~~~~~~+2 {\cal R}_{_{PQ R(M}}H_{_N)}^{^{~R S}} H^{^{PQ}}_{~~_{S}} +\frac{1}{3} ({\nabla}_{_M} H_{_{PQR}})
({\nabla}_{_N} H^{^{PQR}})-({\nabla}_{_P} H_{_{RS M}})
({\nabla}^{^P} H^{^{RS}}_{~~_{N}})\nonumber\\
&&~~~~~~~~~~~~+2 H_{_{MPQ}} H_{_{N R S}} H^{^{T S Q}} H_{_T}^{^{~~R P}}
+2 H_{_{M P Q}} H_{_{NR}} ^{^{~~Q}}   {H^2}^{^{R P}} \Big]+
{\cal O}(\alpha'^2)~=~0,\label{a.2}
\end{eqnarray}
\vspace{-6mm}
\begin{eqnarray}
&&{\nabla}^{^P} H_{_{PMN}} -  ({\nabla}^{^P}\Phi')  H_{_{MNP}}
+\alpha' \Big[{\nabla}^{^P} H^{^{RS}}_{_{~~[M}}{\cal R}_{_{_{{N]} P RS}}} -({\nabla}_{_P} H_{_{R MN}}) {H^2}^{^{PR}}\nonumber\\
&&~~~~~~~~~~~~~~~~~~~~~~~~~~-2 ({\nabla}^{^P}H^{^{Q R}}_{_{~~[M}})H_{_{_{{N]} Q S }}} H_{_{P R}}^{^{~\;S}}\Big]
+{\cal O}(\alpha'^2)~=~0,~~~~~~~\label{a.3}
\end{eqnarray}
\vspace{-6mm}
\begin{eqnarray}
&&2 \Lambda + {\nabla}^2 \Phi' - ({\nabla} \Phi')^2+\frac{2}{3} H^{{2}}
-\alpha' \Big[\frac{1}{4} {\cal R}_{_{MNRS}} {\cal R}^{^{MNRS}}
-\frac{1}{3} ({\nabla}_{_M} H_{_{NRS}})
 ({\nabla}^{^M} H^{^{NRS}})
\nonumber\\
&&~~~~~~~~~~~~~~~~~~~~~~~~~-\frac{1}{2} H^{^{MN}}_{_{~~P}} H^{^{RS P}} {\cal R}_{_{MN RS}}-{\cal R}_{_{MN}} {H^2}^{^{MN}} +\frac{3}{2} H^2_{_{MN}} {H^2}^{^{MN}}\nonumber\\
&&~~~~~~~~~~~~~~~~~~~~~~~~~+\frac{5}{6} H_{_{MN P }} H^{^M}_{_{~~RS}} H^{^{N R}}_{_{~~Q}} H^{^{P S Q}}\Big] +{\cal O}(\alpha'^2)=0,\label{a.4}
\end{eqnarray}
where $H_{_{MNP}}$ defined by $H_{_{MNP}} = 1/2(\partial_{_{M}} B_{_{NP}}
+\partial_{_{N}} B_{_{P M}}+\partial_{_{P}} B_{_{MN}})$ is
the field strength of the field  $B_{_{MN}}$.
We have used the conventional notations $H^2_{_{MN}}=H_{_{MPQ}}  H^{^{PQ}}_{_{~~N}}$, $H^2=H_{_{MNP}}
H^{^{MNP}}$,
${H^2}^{^{MN}} = H^{^{MPQ}} H_{_{PQ}}^{^{~~N}}$ and $({\nabla} \Phi)^2 =\partial_{_{M}} \Phi ~\partial^{^{M}} \Phi$.
${\cal R}_{_{MN}}$ and ${\cal R}_{_{MNPQ}}$ are the Ricci tensor and
Riemann tensor field of the metric
$G_{_{MN}}$, respectively.
Moreover, in Eq. \eqref{a.4}, $\Phi' = \Phi + \alpha' q H^2$ for some coefficient $q$ \cite{c.hull}, and
$\Lambda$ is a cosmological constant. In string theory, the $\Lambda$
is related to the dimension of spacetime, $d$,  and the inverse string
tension, $\alpha'$, whereas in this paper it
is, in some cases,  treated as a free parameter.
We note that round brackets denote the symmetric part on the indicated indices whereas square brackets
denote the antisymmetric part.

On the other hand, the conditions for conformal
invariance (Eqs. \eqref{a.2}-\eqref{a.4}) can be interpreted as field equations for ${G}_{_{MN}}$,
${B}_{_{MN}}$ and $\Phi$ of the string effective action \cite{Metsaev}.
As shown in Ref. \cite{c.hull}, in d=26 (where $\Lambda =0$), the string effective action up to the first
order in  $\alpha'$ is given by
\begin{eqnarray}\label{a.5}
S_{_{eff}} = \int\! d^{^{D}}\hspace{-2mm}X \sqrt{-G} ~ e^{-\Phi}
\hspace{-4mm}&&\hspace{-6mm}\Big\{{\cal R}-\frac{1}{3} H^2  +({\nabla} \Phi)^2 + \alpha' \Big[\frac{1}{4}{\cal R}_{_{MNPQ}} {\cal R}^{^{MNPQ}}
-\frac{1}{2}  {\cal R}_{_{MN P Q}} H^{_{MN S}} H^{^{PQ}}_{{~~_S}}\nonumber\\
&+&\hspace{-2mm}\frac{1}{6}  H_{_{MNP}} H^{^{P Q R}}{H_{_R}}^{^{N S}} {H_{_{S Q}}}^{^M}+ \frac{1}{2}
H^2_{_{MN}} {H^2}^{^{MN}} - {\cal R}_{_{MN}} {\nabla}^{^M} \Phi   {\nabla}^{^N} \Phi   \nonumber\\
&-&\hspace{-2mm}\frac{1}{6}  H^2  ({\nabla} \Phi)^2 + \frac{1}{3} {\nabla}^{^M} \Phi ~ {\nabla}_{_M} H^2
-\frac{1}{6} {\cal R} H^2  + \frac{1}{2} {\cal R} ({\nabla} \Phi)^2  + \frac{1}{2} ({\nabla} \Phi)^2
{\nabla}^2 \Phi\nonumber\\
&-&\hspace{-2mm}{\cal R}_{_{MN}} {\cal R}^{^{MN}}+\frac{1}{36}  (H^2)^2+\frac{1}{4}{\cal R}^2 -\frac{1}{4} (({\nabla} \Phi)^2)^2\Big] \Big\} + {\cal O} (\alpha'^2).
\end{eqnarray}
As announced in the introduction, the G\"{o}del spacetimes can be considered as exact solutions
in string theory for the full ${\cal O}(\alpha')$ action including both dilaton and axion fields \cite{Barrow}.
In Ref. \cite{Barrow}, it has been considered a convenient ansatz for both dilaton and
axion fields in an orthonormal frame.
In the next section, we obtain the other forms of solutions for
the two-loop beta-function equations including the G\"{o}del spacetimes, the field strength $H$
and dilaton $\Phi$ in such a way that we do not work in an orthonormal frame.

\section{\label{Sec.III} G\"{o}del spacetimes  as solutions in string theory for the full ${\cal O}(\alpha')$ action}

\subsection{\label{Sec.III.1} G\"{o}del spacetimes in the cylindrical coordinates $(\hat{t}, \hat{r}, \hat{\varphi}, \hat{z})$}

Among the known exact solutions of Einstein field equations,
the G\"{o}del and G\"{o}del-type metrics \cite{godel} play a special role. It was shown within the usual general relativity
that these solutions describe rotating string cosmologies, and allow for the existence of CTC's.
It is a well-known result that all G\"{o}del-type metrics, i.e., homogeneous spacetimes exhibiting vorticity, characterized by $\Omega$,
and a given value of $m$ parameter can be rewritten in
cylindrical coordinates $(\hat{t}, \hat{r}, \hat{\varphi}, \hat{z})$ as
\begin{eqnarray}
ds^2~=~-d\hat{t}^2 - 2C(\hat{r})~d\hat{t} d \hat{\varphi}+G(\hat{r})~d \hat{\varphi}^2+d\hat{r}^2 +d\hat{z}^2,\label{b.1}
\end{eqnarray}
where the functions $C(\hat{r})$ and  $G(\hat{r})$  must obey the relations
\begin{eqnarray}
C(\hat{r})&=&\frac{4 \Omega}{m^2}~\sinh^2(\frac{m\hat{r}}{2}),\nonumber\\
G(\hat{r})&=&\frac{4}{m^2}~\sinh^2(\frac{m\hat{r}}{2})\Big[1+(1-\frac{4\Omega^2}{m^2})  \sinh^2(\frac{m\hat{r}}{2})\Big].\label{b.2}
\end{eqnarray}
We note that $m^2=2\Omega^2$ is a particular case of the hyperbolic class which corresponds to the original G\"{o}del
solution \cite{godel}. An interesting aspect of G\"{o}del-type solutions is the possibility for existence of CTC's.
The existence of  CTC's, which allows for violation of
causality, depends upon the sign of the metric function $G(\hat{r})$.
Indeed, from Eqs. \eqref{b.1} and \eqref{b.2} one finds that the circles defined by $\hat{t}=t_{_0}, \hat{r} =r_{_0}, \hat{\varphi}\in [0 , 2\pi], \hat{z}=z_{_0}$
become CTC's whenever $G(\hat{r})<0$  \cite{Reboucas}. In fact, the range, $m^2 \geq 4 \Omega ^2$, does not present CTC's.

\subsubsection{Solution up to zeroth order in $\alpha'$}

The equations \eqref{a.2}-\eqref{a.4} up to zeroth order in $\alpha'$ are satisfied only when
we have $m^2 = 4\Omega^2$ with no CTC's.
In this case, the field strength and dilaton field are, respectively, given by
\begin{eqnarray}
H &=&\frac{1}{2} \sinh (m\hat{r}) d\hat{t} \wedge d\hat{r} \wedge d\hat{\varphi},\\
\Phi &=& h \hat{z}+b,\label{b.2.1}
\end{eqnarray}
where $b$ is an arbitrary constant, and  $h^2 = 2 (\Lambda - 2\Omega^2)$.
It is noteworthy that this solution is in agreement with those of Ref. \cite{Barrow}.

\subsubsection{Solutions up to first order in $\alpha'$}
Below we discuss the solutions of equations \eqref{a.2}-\eqref{a.4} up to first order in $\alpha'$
possessing the G\"{o}del spacetimes in a coordinate base $(\hat{t}, \hat{r}, \hat{\varphi}, \hat{z})$.
Our solutions are, in general, classified into two special classes:

\bigskip
{\bf Class~I:}~In this class of solutions the field strength and dilaton field are, respectively, given by
\begin{eqnarray}
H =\mathbb{C} \sinh (m\hat{r}) d\hat{t} \wedge d\hat{r} \wedge d\hat{\varphi},~~~~~~~~~~~\Phi = h \hat{z}+b,\label{b.3}
\end{eqnarray}
for some constants $\mathbb{C}, h, b$.
The equations \eqref{a.2}-\eqref{a.4} together with the
fields given by Eq. \eqref{b.3} possess a G\"{o}del solution with the metric \eqref{b.1}
if the following conditions hold between the constants $m, \Omega, \mathbb{C}, h, \alpha'$  and $\Lambda$:
\smallskip

{\bf (i)}~The first constraint  to satisfy the field equations \eqref{a.2}-\eqref{a.4} with the metric \eqref{b.1} and the fields  \eqref{b.3} is that
\begin{eqnarray}
m^2 = 4\Omega^2,~~ ~\alpha' =\frac{1}{\Omega^2(1-12\mathbb{C}^2)},~~~~h^2 = 2 \Lambda - \Omega^2\frac{(176 \mathbb{C}^4 -56 \mathbb{C}^2 +3)}{(1-12\mathbb{C}^2)}.\label{b.4}
\end{eqnarray}
This confirms that it is possible to obtain a G\"{o}del solution
with no CTC's. By noting the relation \eqref{b.4}, the field strength $H$ depending on $\mathbb{C}$ can vanish. Then, the second relation
of \eqref{b.4} gives the velocity of rotation of the G\"{o}del universe in
terms of the inverse string tension; moreover, there is another constraint, $h^2 = 2 \Lambda - 3\Omega^2$,
which shows that the cosmological term has to be positive. This particular case, $\mathbb{C}=0$, is in agreement with those of Ref. \cite{Barrow}.

\smallskip

{\bf (i$'$)}~ Similar to the case (i) we have $m^2 = 4\Omega^2$ with no CTC's.  In this case, the value of $\mathbb{C}$ is fixed to
be $\mathbb{C}^2= 1/4$, and $h^2 = 2 \Lambda - 4\Omega^2 (1+2 \alpha' \Omega^2)$.

\smallskip

{\bf (ii)}~In this case, the relation between the constants is given by
\begin{eqnarray}
\mathbb{C}^2 =1- \frac{3\Omega^2}{m^2},~~~ \alpha' =\frac{1-\mathbb{C}^2}{\Omega^2(1-10\mathbb{C}^2)},~~~~\frac{h^2}{2} = \Lambda - \Omega^2\frac{(1-4 \mathbb{C}^2)}{(1-\mathbb{C}^2)}.\label{b.5}
\end{eqnarray}
The first relation requires that $m^2 \geq 3 \Omega^2$ for $\mathbb{C}^2 \geq 0$.
The range of $3 \Omega^2 \leq m^2< 4 \Omega^2$ allows CTC's. When $m^2 = 3 \Omega^2$, the field strength vanishes, then,
$\alpha' = 1/ \Omega^2$ and $\Lambda > h^2/2$.

\smallskip

{\bf (iii)}~The last case refers to the following relation between the constants
$m, \Omega, \mathbb{C}, h, \alpha'$ , and $\Lambda$
\begin{eqnarray}
\mathbb{C}^2 =\frac{\Omega^2}{m^2},~~~~~ \alpha' =\frac{\mathbb{C}^2}{\Omega^2(1-6\mathbb{C}^2)},~~~~~~h^2 = 2 \Lambda - \frac{\Omega^2(1-4 \mathbb{C}^2)^2}{\mathbb{C}^2(1-6\mathbb{C}^2)}.\label{b.6}
\end{eqnarray}
From the first relation of \eqref{b.6} one can easily deduce that the
field strength must not vanish if a solution is to exist and act as
a source of rotation.
In this situation by fixing $\mathbb{C}^2$ to $1/2$ we then have the original
G\"{o}del solution\footnote{In this case, one gets $\alpha'=-\frac{1}{4 \Omega^2}$.
In string theory, the resulting spectrum of the bosonic string contains a finite number of massless
and infinitely many massive excitations with ${mass}^2=\frac{n}{\alpha'}$ with $n \in N$.
Among the states there are also tachyons with ${mass}^2<0$ implying that the vacuum is unstable.
This is unavoidable in the bosonic string.} for Eqs. \eqref{a.2}-\eqref{a.4}.

\bigskip

{\bf Class~II:}~The corresponding field strength  to this class of solutions is given by
\begin{eqnarray}
H = \mathbb{C} \sinh (m\hat{r}) d\hat{t} \wedge d\hat{r} \wedge d\hat{\varphi}
+\mathbb{D} \sinh (m\hat{r}) d\hat{r} \wedge d\hat{z} \wedge d\hat{\varphi},\label{b.7}
\end{eqnarray}
for some constants $\mathbb{C}, \mathbb{D}$. In this case, dilaton field is assumed to be constant, $\Phi = b$.
Using these, the equations \eqref{a.2}-\eqref{a.4} with the metric \eqref{b.1} are satisfied
if the following relation hold between the constants $m, \Omega, \mathbb{C}, \mathbb{D}, \alpha'$ and $\Lambda$:
\begin{eqnarray}
\mathbb{C}^2&=&\frac{216 \Omega^4 -(108 m^2+ 48 \Lambda) \Omega^2 +(17m^2 +6\Lambda)m^2+
(m^2-8\Omega^2) \Gamma}{88 m^2 \Omega^2},~~\nonumber\\
\alpha' &=& \frac{(m^2-8\Omega^2)}{2m^2 \Omega^2(11+6\mathbb{C}^2) -3m^4-44 \Omega^4},\nonumber\\
\mathbb{D}^2 &=& \frac{(m^2-4\Omega^2)[3\Omega^2 - m^2(1-\mathbb{C}^2)]}{m^2(m^2 -8\Omega^2)},\label{b.8}
\end{eqnarray}
where $\Gamma = \sqrt{80 \Omega^4 -(104 m^2-160 \Lambda) \Omega^2 +(25 m^2 -60 \Lambda)m^2 +36 \Lambda^2}$.
\smallskip

\subsection{\label{Sec.III.2} G\"{o}del spacetimes in new coordinates $(t, r, \varphi, z)$}

Since our main aim in the present work is to study the T-dualization of the G\"{o}del spacetimes we write down the metric \eqref{b.1} in a simpler form.
To this end, one uses the coordinate transformation
\begin{eqnarray}
\sqrt{\beta}~{t} &=& \Big\{\frac{\sqrt{\beta}}{l} ~ \hat{t}- \hat{\varphi} +2\arctan \big[e^{-m\hat{r}}~ \tan (\frac{\hat{\varphi}}{2})\big]\Big\},
\nonumber\\
{r} &=& \cosh (m\hat{r})+\cos\hat{\varphi} \sinh (m\hat{r}),\nonumber\\
\sqrt{\beta}~r {\varphi} &=& \sin \hat{\varphi} ~\sinh (m\hat{r}),\nonumber\\
{z} &=& \frac{\hat{z}}{l},\label{b.9}
\end{eqnarray}
to obtain
\begin{eqnarray}
ds^2~=~l^2\Big[-d{t}^2 + \frac{dr^2}{r^2}+(\beta-1)r^2~ d{\varphi}^2-2r~ dt d\varphi +d{z}^2\Big],\label{b.10}
\end{eqnarray}
such that the following relations must be held between the constants
${l}, {\beta}, m$ and  $\Omega$:
\begin{eqnarray}
l = \frac{1}{m},~~~{\beta} = \frac{m^2}{4 \Omega^2}.\label{b.11}
\end{eqnarray}
Here, the condition of the existence of CTC's, $m^2 < 4 \Omega^2$, is induced on the range of the parameter $\beta$.
In fact, for the range $\beta \geq 1$ we do not encounter CTC's.
In this case, the original G\"{o}del metric \cite{godel} is recovered when we take $\beta=1/2$.
As it is seen, the metric \eqref{b.10} is a direct product of $\mathbb{R}$ associated with the
coordinate $z$ and the three-dimensional metric of $(t, \varphi, r)$.
For the case of $\beta=1$ one may use the following transformation
\begin{eqnarray}
dt = \rho(d\tau - dx),~~~~ r=\rho,~~~~~~\varphi=x,\label{b.11.1}
\end{eqnarray}
then, the metric becomes
\begin{eqnarray}
ds^2~=~l^2\Big[-\rho^2 d{\tau}^2 + \frac{d\rho^2}{\rho^2}+ \rho^2 dx^2  +d{z}^2\Big],\label{b.11.2}
\end{eqnarray}
so when $\beta=1$, the metric \eqref{b.10} is locally $AdS_3 \times \mathbb{R}$ space.

\subsubsection{Solution up to zeroth order in $\alpha'$}

One can show that only $\beta =1$ case of the metric \eqref{b.10}
satisfies the field equations \eqref{a.2}-\eqref{a.4} up to zeroth order in $\alpha'$.
In this case, a solution including the metric, field strength and dilaton field is given by
\begin{eqnarray}
ds^2&=&l^2\Big[-d{t}^2 + \frac{dr^2}{r^2}-2r~ dt d\varphi +d{z}^2\Big],\nonumber\\
H &=&\frac{l^2}{2}  dt \wedge dr \wedge d{\varphi},\nonumber\\
\Phi &=& f {z}+b,\label{b.11.3}
\end{eqnarray}
where $f^2=-1+ 2 l^2 \Lambda $.

\subsubsection{Solutions up to first order in $\alpha'$}

In order to obtain new solutions up to first order in $\alpha'$ we now solve the field
equations \eqref{a.2}-\eqref{a.4} for the metric \eqref{b.10}.
Here, the forms of our solutions including
the field strength and dilaton field are given by two special classes A and B:

\bigskip
{\bf Class~A:}~In this class of solutions the field strength $H$ and dilaton field $\Phi$ are given by
\begin{eqnarray}
H = \mathbb{E}~ dt \wedge dr \wedge d{\varphi},~~~~~~~~~~~\Phi = f {z}+b,\label{b.12}
\end{eqnarray}
for some constants $\mathbb{E}, f, b$.
The field equations \eqref{a.2}-\eqref{a.4} are then satisfied with the metric \eqref{b.10}
together with the fields \eqref{b.12}
if the following four conditions held between the constants $l, \beta, \mathbb{E}, f, \alpha'$ and $\Lambda$:
\smallskip

{\bf (1)}~The first condition is devoted to a special condition on the parameter $\beta$, and that is $\beta=1$.
The rest of constants are then related to each other in the following way:
\begin{eqnarray}
\alpha' =\frac{4 l^6}{l^4-12 \mathbb{E}^2},~~~~~~~~f^2 = \frac{176 \mathbb{E}^4 +8l^4 (12 l^2 \Lambda -7) \mathbb{E}^2 +l^8 (3-8l^2 \Lambda)}{4l^4(12 \mathbb{E}^2-l^4)}.\label{b.13}
\end{eqnarray}
This result confirms that it is possible to have a solution corresponding to zero field strength, that is, $\mathbb{E}=0$.
Then, it is followed that $\alpha' =4 l^2$ and $f^2 = 2 l^2 \Lambda -3/4$.

\smallskip

{\bf (1$'$)}~In addition to the first condition, we have another solution corresponding to  $\beta=1$ and nonzero field strength
$\mathbb{E}^2=l^4/4$, so that the relation between other constants may be expressed as $f^2=-1+ 2 l^2 \Lambda - \alpha'/(2l^2)$.

\smallskip

{\bf (2)}~ The field equations \eqref{a.2} and \eqref{a.3} are also fulfilled if the values of $\mathbb{E}$ and $\alpha'$
can now be expressed in terms of the parameter $\beta$
\begin{eqnarray}
\mathbb{\mathbb{E}}^2 = \frac{l^4}{4} (4 \beta -3),~~~~~~~~~\alpha' =\frac{2 \beta l^2}{5-6 \beta}.\label{b.14}
\end{eqnarray}
This is an interesting case in which the field strength depends on the parameter $\beta$ of the metric.
From the relation \eqref{b.14} one can easily deduce that $\beta \in [3/4 , +\infty)-\{5/6\}$.
We note that in this case, the range of $\beta \in [3/4 , 1)-\{5/6\}$ allows CTC's.
The relation between other constants can be obtained from Eq. \eqref{a.4}, which gives, $f^2=2(1+ l^2 \Lambda - 1/\beta)$.

\smallskip

{\bf (3)}~~The last condition refers to a G\"{o}del solution with a nonzero field strength,
$\mathbb{E}^2=l^4/4$.  From Eqs. \eqref{a.2} and \eqref{a.3} the value of $\alpha'$ is expressed in terms of the parameter $\beta$, obtaining
\begin{eqnarray}
\alpha' =\frac{2 \beta l^2}{2\beta -3}.\label{b.15}
\end{eqnarray}
Finally after using Eq. \eqref{a.4}, we obtain the relation
\begin{eqnarray}
\frac{f^2}{2}= \frac{\beta^2 (2 l^2 \Lambda -1)+\beta (2- 3 l^2 \Lambda)-1}{\beta(2\beta -3)}.\label{b.16}
\end{eqnarray}
This is only case of the class A which is valid for $\beta=1/2$.
 Putting $\beta=1/2$, we then get $\alpha'=-l^2/2$. Also, from Eq.  \eqref{b.16} it is followed that $\Lambda \geq -1/(4 l^2)$,
which shows that the cosmological term can be considered to be negative.

\bigskip
{\bf Class~B:}~ Class B of solutions is devoted to a constant dilaton field $\Phi=b$ and
the field strength
\begin{eqnarray}
H = \mathbb{E}~ dt \wedge dr \wedge d{\varphi}+ \mathbb{N} dr \wedge d{\varphi} \wedge dz,\label{b.17}
\end{eqnarray}
for some constants $\mathbb{E}, \mathbb{N}$, together with the metric \eqref{b.10}. The equations of motion \eqref{a.2}-\eqref{a.4} are then fulfilled in this
general case. From Eqs. \eqref{a.2} and \eqref{a.3} we obtain
\begin{eqnarray}
\alpha' &=& \frac{4\beta l^6 (\beta -2)}{l^4(12 \beta^2 -22 \beta +11)-12 \mathbb{E}^2},\nonumber\\
\mathbb{N}^2 &=& \frac{(\beta -1)[4 \mathbb{E}^2+l^4(3-4\beta)]}{4(\beta -2)}.\label{b.18}
\end{eqnarray}
Finally, Eq. \eqref{a.4} is satisfied if
{\small \begin{eqnarray}
\mathbb{E}^2 = \frac{l^4}{44}\big[34 \beta^2 -54 \beta +12 \beta l^2 \Lambda (\beta -2) +27+ 2(\beta-2) \Xi\big],\label{b.19}
\end{eqnarray}}
where $\Xi = \sqrt{\beta^2(5-6l^2 \Lambda)^2+ 2\beta(20 l^2 \Lambda-13) +5}$.
We have thus obtained some new solutions for the field equations of
bosonic string effective action up to the first order in $\alpha'$ in the forms of classes A and B.
These solutions will be useful in studying the T-dualization of G\"{o}del string cosmologies.
We shall address this problem in Secs. \ref{Sec.V}, \ref{Sec.VI} and \ref{Sec.VII}.

\section{\label{Sec.IV} A short review of PL T-duality with spectators}

In this section, we recall the main features of PL T-duality transformations at the level of
the $\sigma$-model. For the description of PL T-duality, we need to
introduce the Drinfeld double group $ D$ \cite{Drinfeld}, which by definition has a pair of
maximally isotropic subgroups $G$ and ${\tilde {G}}$
corresponding to the subalgebras ${\cal G}$ and ${\tilde {\cal G}}$, respectively.
The generators of  ${\cal G}$ and ${\cal \tilde G}$ are denoted,
respectively, $T_a$ and ${\tilde T}^a$, $a=1, . . . , dim { G}$ and
satisfy the commutation relations
\begin{eqnarray}\label{c.1}
[T_a , T_b] &=& {f^c}_{ab} ~T_c,~~~
[{\tilde T}^a , {\tilde T}^b] = {{\tilde f}^{ab}}_{\; \; \: c} ~{\tilde T}^c,\nonumber\\
{[T_a , {\tilde T}^b]} &=& {{{\tilde f}^{bc}}_{\; \; \; \:a} {T}_c + {f^b}_{ca} ~{\tilde T}^c}.
\end{eqnarray}
The Lie algebra structure defined by \eqref{c.1} is called the Drinfeld double ${\cal D}$. The structure constants
${f^c}_{ab}$ and ${{\tilde f}^{ab}}_{\; \; \: c}$ are subject to the Jacobi identities and the following mixed Jacobi identities
\begin{eqnarray}\label{c.2}
{f^a}_{bc}{\tilde{f}^{de}}_{\; \; \; \; a}=
{f^d}_{ac}{\tilde{f}^{ae}}_{\; \; \; \;  b} +
{f^e}_{ba}{\tilde{f}^{da}}_{\; \; \; \;  c}+
{f^d}_{ba}{\tilde{f}^{ae}}_{\; \; \; \; c}+
{f^e}_{ac}{\tilde{f}^{da}}_{\; \; \; \; b}.
\end{eqnarray}
In addition, the Drinfeld double ${\cal D}$ is equipped with an invariant inner product
$<.~ ,~ .>$ with the following properties
\begin{eqnarray}\label{c.3}
<T_a , {\tilde T}^b> &=& {\delta _a}^{~b},\nonumber\\
<T_a , T_b> &=& <{\tilde T}^a, {\tilde T}^b> ~ =~ 0.
\end{eqnarray}
Let us now consider a $d$-dimensional manifold ${\cal M}$ and some coordinates $X^{^{M}} = (x^{\mu} , y^i)$ on it,
where $x^\mu,~ \mu = 1, . . . , dim  G$ stand for the coordinates of Lie group $G$ acting freely from right on ${\cal M}$ and
$y^i,~i = 1, . . . , d-dim { G}$ are
the coordinates labeling the orbit $O$ of $ G$ in the target space ${\cal M}$.
We note that  the coordinates $y^i$ do not participate in the PL T-duality transformations
and are therefore called spectators \cite{Sfetsos1}.

Before proceeding to write the $\sigma$-models let us introduce the components of the right-invariant Maurer-Cartan
forms $(g^{-1} \partial_{\alpha} g)^a \equiv R_{\alpha}^a=\partial_{\alpha} x^{\mu} R_{\mu}^{~a}$,
where $g$ is an element of the Lie group ${G}$ corresponding to the Lie algebra ${\cal G}$.
For notational convenience we will also use  $R_{\alpha}^i=\partial_{\alpha} y^i$.
In order to define $\sigma$-models with PL T-duality symmetry,
it is convenient to define matrices $a(g)$, $b(g)$ and $\Pi(g)$ in the following way
\begin{eqnarray}
g^{-1} T_{{_a}}~ g &=& a_{_{a}}^{^{~b}}(g) ~ T_{{_b}},\nonumber\\
g^{-1} {\tilde T}^{{^a}} g &=&
b^{^{ab}}(g)~ T_{{_b}}+(a^{-1})_{_{b}}^{^{~a}}(g)~{\tilde T}^{{^b}},\nonumber\\
\Pi^{^{ab}}(g) &=& b^{^{ac}}(g)~ (a^{-1})_{_{c}}^{^{~b}}(g).\label{c.4}
\end{eqnarray}
Then, the original $\sigma$-model is defined by the action \cite{{Klim1},{Klim2},{Sfetsos1}}
\begin{eqnarray}\label{c.5}
S = \frac{1}{2} \int d\sigma^{+}  d\sigma^{-}&& \hspace{-2mm}\Big[{{E}_{_{ab}}}~
R_{+}^a \;R_{-}^b + \phi^{{(1)}}_{a j} R_{+}^a \partial_{-} y^{j}+
\phi^{{(2)}}_{i b}  \partial_{+} y^{i} R_{-}^b  +\phi_{_{ij}}
\partial_{+} y^{i} \partial_{-} y^{j} \Big]\nonumber\\
&&~~~~~~~~~~~~~~~~~~~~~~~~~~~~~~~~~~~~~~-\frac{1}{4 \pi} \int d\sigma^{+}  d\sigma^{-} ~R^{^{(2)}} \Phi(X).
\end{eqnarray}
Here, we have used the standard light-cone variables on the worldsheet,
$\sigma^{\pm} =(\tau \pm \sigma)/2$ together with $\partial_{\pm}=\partial_{\tau} \pm \partial_{\sigma}$.
The backgrounds appearing in this action are given in matrix notation by \cite{{Sfetsos1}}
\begin{eqnarray}
{{E}}&=&\big(E^{{-1}}_{0}+ \Pi\big)^{-1},\label{c.6}\\
\phi^{{(1)}}&=& {{E}}~E^{{-1}}_{0}~F^{^{(1)}},\label{c.7}\\
\phi^{{(2)}}&=& F^{^{(2)}}~ E^{{-1}}_{0}~{{E}},\label{c.8}\\
\phi&=& F -F^{^{(2)}}~\Pi~{{E}}~E^{{-1}}_{0}~F^{^{(1)}}.\label{c.9}
\end{eqnarray}
The matrices $(E_{0}, F, F^{^{(1)}}, F^{^{(2)}})$
are all functions of the variables $y^{i}$ only.

Similarly we consider another $\sigma$-model for the $d$ field variables ${\tilde X}^{^{M}} = ({\tilde x}^{\mu} , y^i)$,
where ${\tilde x}^{\mu},~ \mu = 1, 2, . . . , dim \tilde G$
parameterize an element $\tilde g$ of a Lie group ${\tilde {G}}$, whose dimension is, however,
equal to that of $G$. The rest of the variables are the same $y^i$ used in \eqref{c.5}.
Accordingly, we introduce the components of the right-invariant Maurer-Cartan forms
$(\tilde g^{-1} \partial_{\pm} \tilde g)_a={\tilde R}_{{\pm}_a}=\partial_{\pm} {\tilde x}^{\mu} {\tilde R}_{\mu a}$ on the Lie group $\tilde G$.
The corresponding $\sigma$-model has the form
\begin{eqnarray}\label{c.11}
\tilde S &=& \frac{1}{2} \int d\sigma^{+}  d\sigma^{-}\Big[{{{\tilde E}}^{{ab}}}~
{\tilde R}_{+_{a}}{\tilde R}_{-_{b}}+{\tilde \phi}^{\hspace{0mm}{(1)^{ a}}}_{~~~j} ~{\tilde R}_{+_{a}}\partial_{-} y^{j}+
{\tilde \phi}^{\hspace{0mm}{(2)^{ b}}}_{i} \partial_{+} y^{i} ~{\tilde R}_{-_{b}}
+{\tilde \phi}_{_{ij}} \partial_{+} y^{i} \partial_{-} y^{j}\Big]\nonumber\\
&&~~~~~~~~~~~~~~~~~~~~~~~~~~~~~~~~~~~~~~~~~~
~~~~~~~~~~~~~~~~-\frac{1}{4 \pi} \int d\sigma^{+}  d\sigma^{-} ~R^{^{(2)}}  {\tilde \Phi({\tilde X})}.
\end{eqnarray}
The backgrounds of the dual theory are related to those of the original one by \cite{{Klim1},{Sfetsos1}}
\begin{eqnarray}
{{\tilde E}} &=& \big(E_{0}+ {\tilde \Pi}\big)^{-1},\label{c.12}\\
{\tilde \phi}^{{(1)}} &=&  {{\tilde E}}~F^{^{(1)}},\label{c.13}\\
{\tilde \phi}^{{(2)}} &=& - F^{^{(2)}} ~{{\tilde E}},\label{c.14}\\
{\tilde \phi}           &=& F-F^{^{(2)}} ~{{\tilde E}}~F^{^{(1)}},\label{c.15}
\end{eqnarray}
where ${\tilde \Pi}$ is defined as in \eqref{c.4} by replacing untilded quantities with
tilded ones.

Notice that the duality transformation must be supplemented by a
correction that comes from integrating out the fields on the dual group
in path integral formulation in such a way that it can be absorbed at the one-loop level into the transformation of the dilaton field.
The correct formula of dilaton transformation is, for the PL T-duality case,  given by \cite{vonUnge}
\begin{eqnarray}
\Phi &=&\phi^{^{(0)}} +\log |\det {{E}}| - \log |\det {E_0}|  - \log |\det {a(g)}|,\label{c.16.1}\\
{\tilde \Phi} &=& \phi^{^{(0)}}+\log |\det {{\tilde E}}| - \log |\det {{\tilde a} ({\tilde g})}|,\label{c.16.2}
\end{eqnarray}
where $\phi^{^{(0)}}$ is the dilaton that makes the original $\sigma$-model conformal (up to the one-loop order)
and may depend on both group and spectator coordinates. Accordingly,
the dual background can also be conformal at the one-loop level with a new dilaton field obtaining from
equation \eqref{c.16.2}.
The transformations \eqref{c.16.1} and \eqref{c.16.2} were based on a regularization of a functional
determinant in the path integral formulation of PL duality \cite{vonUnge} (see also \cite{Tyurin}).
PL T-duality is a canonical transformation and two $\sigma$-models related by PL
duality are, therefore, equivalent at the classical level \cite{Sfetsos6}.
It has been shown that \cite{N.Mohammedi} relations \eqref{c.16.1} and \eqref{c.16.2} only hold at the one-loop level
for both $\sigma$-models admitting PL duality if the traces of the structure constants
of each Lie algebra constituting the Drinfeld double are zero.
Equivalence can hold in some
special cases but it fails in most cases. In this respect,
checking the equivalence by studying conformal invariance
(the vanishing of the beta-function equations) is important.
But, since after a classical canonical transformation, the
equivalence always holds up to first order in Planck's
constant in the semiclassical expansion (corresponding
to one-loop order in $\sigma$-model language), only the two-loop
order is the first real test of quantum equivalence
of the two different $\sigma$-models related by PL T-duality \cite{{Wulff22},{Hassler2}}.
As mentioned in the introduction section, in the case of  Abelian T-duality,
the two-loop $\sigma$-model corrections  were obtained by KM in \cite{KM}.
There, one can find the duality transformation of the dilaton field at two-loop level.

Let us turn into the actions \eqref{c.5} and \eqref{c.11}. These actions correspond to PL T-dual $\sigma$-models \cite{{Klim1},{Klim2}}.
If the group $G(\tilde G)$ besides having free action on the manifold ${\cal M}( \tilde {\cal M})$,
acts transitively on it, then the corresponding manifold ${\cal M}( \tilde {\cal M})$ will be the same as the group $G(\tilde G)$.
In this case only the first term appears in the actions \eqref{c.5} and \eqref{c.11}.
The T-duality transformations are said to be Abelian \cite{{Buscher1},{Buscher2}}
or non-Abelian  \cite{{Ossa},{Alvarez}} according to the nature of the Lie algebra
formed by the generators of the isometries.
Notice that:

$\bullet$~ If both the Lie groups $G$ and $\tilde G$ become the isometry groups of the manifolds ${\cal M}$ and $\tilde {\cal M}$, respectively,
namely, both $G$ and $\tilde G$ are chosen to be Abelian groups $({f^a}_{bc} = {\tilde{f}^{ab}}_{\; \; \; c} =0)$,
then we get  ${\Pi}(g)={\tilde \Pi}(\tilde g)=0$, recovering thus the standard Abelian duality.
In this case, the dilaton transformation \eqref{c.16.1} turns into $\Phi  = \phi^{^{(0)}}$.
By remembering that $\phi^{^{(0)}}$ makes the original model conformal at one-loop level we obtain the final result
for the dual dilaton, giving
\begin{eqnarray}
{\tilde \Phi} = \phi^{^{(0)}}- \log |\det {E_0}|.\label{c.16.3}
\end{eqnarray}
Indeed, this new dilaton field  guaranties the conformality of dual background up to the one-loop order.

$\bullet$~ In case of non-Abelian T-duality the former represents
group of symmetries of the original $\sigma$-model, while the latter is
Abelian  $({\tilde{f}^{ab}}_{\; \; \; c} =0)$.
 Furthermore, there are  Drinfeld doubles where both $G$ and $\tilde G$ are non-Abelian.
In such a case the symmetry of the original model is replaced with the
so-called PL symmetry (or generalized symmetry), and the
full PL T-duality transformation \cite{{Klim1},{Klim2}} applies.
We now wish to apply the above discussions to study the Abelian T-dualization of the G\"{o}del string cosmologies
in the next section.


\section{\label{Sec.V} Abelian T-dualization of G\"{o}del string cosmologies}

In order to study Abelian T-duality in Buscher's construction \cite{{Buscher1},{Buscher2}} one starts with a manifold ${\cal M}$
with metric ${G}_{_{MN}}$,
antisymmetric tensor ${B}_{_{MN}}$ and dilaton field $\Phi$, and requires the metric to admit at least one
continuous Abelian isometry leaving invariant the action of $\sigma$-model constructed out of $(G, B, \Phi)$.
As announced in the introduction, PL T-duality proposed by Klim\v{c}ik and \v{S}evera
is a generalization of Abelian and non-Abelian T-dualities. It was then shown that Buscher's duality transformations
can be obtained from  the PL T-duality approach by a convenient choice of
the spectator-dependent matrices \cite{Klim4} (see also \cite{EMR13}).
In this section, we obtain all possible Abelian T-duals of the G\"{o}del string cosmologies by using the approach of
PL T-duality in the presence of spectators. In this regard, the Lie groups $ G$ and $\tilde { G}$ acting freely on the target space manifolds
${{\cal M}} \approx O \times G$ and ${\tilde {\cal M}} \approx O \times \tilde G$, respectively,
can be considered to be Abelian Lie groups of dimensions one, two
and three. In all cases, both structures $\Pi(g)$ and ${\tilde \Pi}{(\tilde g)}$ are vanished as
mentioned above.

When the dualizing is implemented by the shift of the directions $t$, $z$ and $(t, z)$ we show that the pair of the mutually dual models can be
conformally invariant at the one-loop level, in a way that the corresponding dual dilaton field is found by using transformation
\eqref{c.16.3}.
It is shown that for the duals obtained through the dualizing on these directions without further corrections, we are still able to obtain two-loop
solutions.
On the other hand, the dual models obtained by the rest of the directions, $\varphi$, $(t, \varphi)$, $(\varphi, z)$ and $(t, \varphi, z)$,
don't remain conformally invariant at the one-loop level; we just check their two-loop conformal invariance conditions.

\subsection{Abelian T-duality with one-dimensional Abelian Lie group}

\subsubsection{{Dualizing with respect to the coordinate $\varphi$}}
\label{sec.5.1.1}

Here we assume that target space ${{\cal M}} \approx O \times  G$ is defined by coordinates $X^{{^M}}=(x; r,t,z)$.
The coordinates of orbit $O$
are represented by $y^i=(r,t,z)$, while $x$ is the coordinate of one-dimensional Abelian Lie group
$G=A_1$ parameterizing by element $g~=~e^{x T}$.
In order to write the original $\sigma$-model on the manifold ${\cal M}$ we need to determine the model couplings.
Let us choose the spectator-dependent matrices in the following form
\begin{eqnarray}\nonumber
~~~~~~~~E_{0\;_{ab}}={{\cal C}_{_0}}r^2,~~~~~~~~~~~~~~~~~~~~~~~~~~F^{^{(1)}}_{_{a j}}=\left( \begin{array}{ccc}
	F^{^{(1)}}_{_{1 \bar{1}}}  ~ &   F^{^{(1)}}_{_{1 \bar{2}}}	&	 ~f_{_3}(r,t,z)
	\end{array} \right),~~~~~
\end{eqnarray}
{\small \begin{eqnarray}\label{V.1}
	F_{_{ij}}=\left( \begin{array}{ccc}
\frac{l^2}{r^2}  & f_{_1}(r,t,z)  & f_{_2}(r,t,z) \\
-f_{_1}(r,t,z)      & -{l^2}      & F_{_{23}}	\\
-f_{_2}(r,t,z)      & -F_{_{23}}    & {l^2}	\\
\end{array} \right),~~~~~~~~~~~~~
	F^{{(2)}}_{_{i b}}=-\left( \begin{array}{ccc}
	F^{{(1)}}_{_{1 \bar{1}}}  \\
	F^{{(1)}}_{_{1 \bar{2}}}+2\gamma_{_0} l^2 ~r\\
	f_{_3}(r,t,z)
	\end{array} \right).~~~~~~~
	\end{eqnarray}}
where ${{\cal C}_{_0}}$ is a nonzero constant and
\begin{eqnarray}
\gamma_{_0}^2&=&\frac{{{\cal C}_{_0}}}{l^2(\beta-1)},\nonumber\\
F^{{(1)}}_{_{1 \bar{1}}} &=& h_{_1}(r)+ \int {\Big[\partial_r {f_{_3}(r,t,z)}-2\mathbb{N}\gamma_{_0} \Big] dz}
+\int \Big[l^2\gamma_{_0}(1-{\frac{2\mathbb{E}}{l^2}}) +\partial_r {h_{_2}(r,t)}\Big] dt,\nonumber\\
F^{{(1)}}_{_{1 \bar{2}}} &=& h_{_2}(r,t)+ \int {\partial_t {f_{_3}(r,t,z)} dz},\nonumber\\
F_{_{23}} &=& h_{_3}(t,z)+ \int {\Big[\partial_t {f_{_2}(r,t,z)}-\partial_z {f_{_1}(r,t,z)}\Big] dr},\label{V.2}
\end{eqnarray}
in which $h_i$'s and $f_i$'s are some arbitrary functions
that may depend on some coordinates $r$, $t$ and $z$.
Henceforth, the constants $\mathbb{E}, \mathbb{N}, l$ and $\beta$ are the same ones introduced in Sec. \ref{Sec.III}.
Inserting \eqref{V.1} into Eqs. \eqref{c.6}-\eqref{c.9}
and then employing \eqref{c.5}, the original $\sigma$-model on the manifold $\cal M$ can be derived.
Then corresponding metric and antisymmetric tensor may be expressed as
\begin{eqnarray}
ds^2&=&l^2\Big[-d{t}^2 + \frac{dr^2}{r^2}+ (\beta-1) \gamma_{_0}^2 r^2~ d{x}^2   -2r \gamma_{_0}~ dt dx +d{z}^2\Big],\label{V.3}\\
B &=& f_{_1}(r,t,z) dr \wedge dt+ f_{_2}(r,t,z) dr \wedge dz+F_{_{23}} dt \wedge dz
+ {f_{_3}(r,t,z)} dx \wedge dz\nonumber \\
&&~~~~~~~~~~~~~~~~~~~~~~~~~~~~~~~~~~~~~~~~~~+F^{{(1)}}_{_{1 \bar{1}}}
dx \wedge dr   +\big(\gamma_{_0} l^2 r +F^{{(1)}}_{_{1 \bar{2}}}\big) dx \wedge dt \Big].\label{V.4}
\end{eqnarray}
Carrying out the coordinate transformation
$\varphi =\gamma_{_0} x$ we see that \eqref{V.3} is nothing but the G\"{o}del metric \eqref{b.10}.
Furthermore, one concludes that the field strength corresponding to the $B$-field \eqref{V.4} is the same as \eqref{b.17}.
We note the fact that in the context of PL T-duality,
the matrix $E_0$ must be invertible. With regard to the first relation
of \eqref{V.2} the solution given by \eqref{V.3} and \eqref{V.4} is valid for all values of $\beta$ except for $1$.  Therefore, the original model
don't remain conformally invariant at one-loop level as mentioned in Sec. \ref{Sec.III}.
Finally, one can just verify the field equations \eqref{a.2}-\eqref{a.4}
for the metric \eqref{V.3} and the $B$-field \eqref{V.4} with a constant dilaton field.

The dual background is  obtained from a $\sigma$-model which is constructed on  $1+3$-dimensional manifold
$\tilde {{\cal M}} \approx O \times \tilde { G}$
where ${\tilde G}$ is the same as the $A_1$ Lie group.
Since duality is performed on the $x$, we parameterize ${\tilde G}$ by the coordinate $\tilde x$.
Before proceeding to construct the dual background, let us consider a simpler
form of the spectator-dependent matrices \eqref{V.1}. Indeed, one may take $h_1=h_3=f_1=f_2=0, h_2= (2\mathbb{E}-l^2)r, f_3=2\mathbb{N}r$
and ${{\cal C}_{_0}} = l^2(\beta-1)$. Imposing these conditions on \eqref{V.1}
and then utilizing Eqs. \eqref{c.12}-\eqref{c.15} and also \eqref{c.11},
the metric and antisymmetric tensor field $\tilde B$ of the dual model can be cast in the forms
\begin{eqnarray}
{d\tilde {s}}^2 &=& \frac{l^2 dr^2 }{r^2}+ \frac{1}{(\beta -1)}
\Big[(\frac{4 \mathbb{E}^2}{l^2}-\beta  l^2) dt^2 +\frac{d\tilde{x}^2}{ l^2 r^2} +\big(\frac{4 \mathbb{N}^2}{ l^2}+l^2 (\beta -1)\big)dz^2
\nonumber\\
&&~~~~~~~~~~~~~~~~~~~~~~~~~~~~~~~~~~~~~~~~~~
+\frac{8\mathbb{E} \mathbb{N}}{l^2}  dt dz+\frac{4\mathbb{E}}{l^2 r} d\tilde{x} dt+\frac{4\mathbb{N}}{l^2 r} d\tilde{x} dz\Big],\label{V.6}\\
{\tilde B} &=&\frac{1}{(\beta -1)}  \Big(\frac{1}{r} dt \wedge  d\tilde{x}+2 \mathbb{N} dt\wedge dz\Big).\label{V.7}
\end{eqnarray}
One quickly deduces that the only nonzero component of the field strength corresponding to the $\tilde B$-field \eqref{V.7} is
\begin{eqnarray}
{\tilde H}_{_{{{\tilde x}} t r}}=\frac{1}{2 r^2(\beta-1)}.\label{V.8}
\end{eqnarray}
Let us now clarify the spacetime structure and conformal invariance conditions of the dual model.
We see that the dual background can't, for the $\beta \neq 1$ case, satisfy the field
equations \eqref{a.2}-\eqref{a.4} up to zeroth order in $\alpha'$ (one-loop order).
Accordingly, the dual dilaton don't obey
transformation \eqref{c.16.3}.  Below, we just check the two-loop conformal invariance conditions
of the dual background.
When $\mathbb{N}=0$,  the metric \eqref{V.6} and the field strength
\eqref{V.8} satisfy the field equations \eqref{a.2}-\eqref{a.4} with a dilaton field
in the form of ${\tilde \Phi}={\tilde f} z +b$
provided that the constants $\mathbb{E}, l, {{\tilde f}}, \alpha'$  and $\beta$ satisfy the following relations
\begin{eqnarray}
\mathbb{E}^2=\frac{l^4}{4},~~~~~~~\alpha' = \frac{2 l^2 \beta}{2 \beta -3}, ~~~~~~~~
{\tilde f}^2 = \frac{2 {\beta}^2 (2{\tilde \Lambda} l^2 -1) -2 \beta (3{\tilde \Lambda} l^2 -2)-2}{\beta(2 \beta -3)}.\label{V.9}
\end{eqnarray}
Imposing conditions $\mathbb{N}=0$ and $ \mathbb{E}^2={l^4}/{4}$ on the metric \eqref{V.6} and then performing the coordinate transformation
$r\rightarrow 1/r,~ {{\tilde x}}\rightarrow -l^2(\beta -1) \varphi$
one concludes that it is nothing but the G\"{o}del metric \eqref{b.10} \footnote{
We recall that in the Abelian case with $U(1)$ duality group and another $U(1)$
co-duality group, the target space is a one-dimensional
circle; in other words, we have here the standard $R \rightarrow 1/R$ duality \cite{{Buscher1},{Buscher2}}.}.
This means that the model is self-dual.
Also when $\mathbb{N}=0$ we have another solution in which relationship between the constants are given by
\begin{eqnarray}
\mathbb{E}^2=\frac{l^4(4 \beta -1)}{12},~~~~~~~\alpha' = \frac{6 l^2 \beta}{2 \beta -5}, ~~~~~~~~
{\tilde f}^2 = \frac{2}{3\beta} \big(1+ {\beta} (3{\tilde \Lambda} l^2 -1)\big).\label{V.10}
\end{eqnarray}
In addition to the above, for the metric \eqref{V.6} and the field strength
\eqref{V.8} one verifies the field equations \eqref{a.2}-\eqref{a.4} with a constant dilaton field when $\mathbb{N}$ differs from zero.
In this way we must have
\begin{eqnarray}
\mathbb{E}^2&=&\frac{l^4}{4}(2 \beta^2 -2 \beta +1),~~~~~~~\mathbb{N}^2=\frac{l^4}{2}(\beta -1)^2,~~~~~~~\alpha' = \frac{2 l^2 \beta}{2 -3\beta },\nonumber\\
{\tilde \Lambda} &=& \frac{(\beta -1)^2}{4 l^2 \beta (2- 3{\beta})}.\label{V.11}
\end{eqnarray}

\subsubsection{{Dualizing with respect to the coordinate $t$}}
\label{sec.5.1.2}

In this case the coordinates of the target manifold $\cal M$ are denoted by $(t; r,\varphi,z)$ wherein
$t$ is the coordinate of the Lie group $G$ which the duality is performed on,
while $(r,\varphi,z)$ are the coordinate representations of the orbit $O$.
If we choose the coupling matrices as
{\small \begin{eqnarray}\label{VI.12}
F^{^{(1)}}_{a j}=\left( \begin{array}{ccc}
                    0~& -r(2\mathbb{E}+l^2) ~&0
                      \end{array} \right), F^{^{(2)}}_{i b}=\left( \begin{array}{c}
                    0  \\
                    r(2\mathbb{E}-l^2) \\
                    0
                      \end{array} \right), F_{ij} = \left( \begin{array}{ccc}
                    \frac{l^2}{r^2}  & 0 & 0\\
                       0 & (\beta-1)l^2 r^2 & 2\mathbb{N} r\\
                       0 & -2\mathbb{N} r & l^2
                      \end{array} \right),
\end{eqnarray}}
and $E_{0\;ab}=-l^2$, then using the formulae \eqref{c.6}-\eqref{c.9}
and also \eqref{c.5} one finds that the metric and field strength corresponding to the original $\sigma$-model
take the same form as \eqref{b.10} and \eqref{b.17}, respectively.
According to the results of Sec. \ref{Sec.III}, the
$\beta =1$ case of the G\"{o}del spacetimes with $\mathbb{E}^2=l^4/4$ and $\mathbb{N}=0$ satisfy the one-loop beta-function equations.
In this regard, the dilaton field that makes the original $\sigma$-model conformal
is found to be $\Phi = \phi^{^{(0)}}$. However, according to Sec. \ref{Sec.III} since we want the total dilaton to be ${\Phi} = f z+b$ we need to
choose $\phi^{^{(0)}} =f z+b$ in which $f^2=-1+ 2 l^2 \Lambda $.

It is then straightforward to compute the corresponding dual spacetime. They are read
\begin{eqnarray}
{d\tilde {s}}^2 &=& \frac{l^2 dr^2 }{r^2}+l^2 d z^2 - \frac{1}{l^2} d\tilde{x}^2 + (\beta l^2 -\frac{4\mathbb{E}^2}{l^2}) r^2 d \varphi^2 +
\frac{4 \mathbb{E}}{l^2} r  d\tilde{x} d \varphi,\nonumber\\
{\tilde H} &=&-\frac{1}{2}  dr \wedge d \varphi \wedge d\tilde{x} + \mathbb{N}  dr \wedge d \varphi \wedge d z,\label{V.13}
\end{eqnarray}
where $\tilde x$ is the dualized coordinate of the dual manifold.
We note that the background \eqref{V.13},
unlike the dual background in the preceding case, can be even conformally invariant up
to the zeroth order in $\alpha'$.
For the  $\beta=1$ case of this background the dual dilaton is found by using equation \eqref{c.16.3} to be ${\tilde \Phi} = \phi^{^{(0)}}-\log l^2$.
Finally we get the dilaton by remembering that $\phi^{^{(0)}}= f z+b$ which gives the final result
\begin{eqnarray}
{\tilde \Phi} =  f z+b-\log l^2.\label{V.13.1}
\end{eqnarray}
In addition, the case of $\beta=1$  can satisfy
the field equations \eqref{a.2}-\eqref{a.4} with the same dilaton field \eqref{V.13.1} provided that
\begin{eqnarray}
\mathbb{E}^2=\frac{l^4}{4},~~~~~~~\mathbb{N} = 0, ~~~~~~~~
{f}^2 = -\frac{\alpha'}{2 l^2} +2 l^2 {\Lambda} -1,\label{V.14}
\end{eqnarray}
for each $\alpha'$.  Actually, by imposing the conditions $2\mathbb{E}=-l^2$ and $\mathbb{N}=0$
on the dual background  \eqref{V.13} and then by making use of the transformation
$\tilde{x}\rightarrow l^2 t$, we obtain the same
forms as \eqref{b.10} and \eqref{b.12}, that is, the model is self-dual.

We can also show that the  dual background is conformally invariant up
to first order in $\alpha'$ when $\mathbb{N}$ differs from zero. In this way, if we consider
\begin{eqnarray}
\mathbb{E}^2 &=& \frac{l^4}{4},~~~~~~~~~~\mathbb{N}^2=\frac{l^4 (\beta -1)^2}{2 -\beta},~~~~~~~~~~
\alpha' = \frac{2 l^2 \beta (2-\beta)}{6\beta^2 -11\beta +4},\nonumber\\
{\tilde \Lambda} &=&  \frac{1}{ \beta l^2} \Big(\frac{-6\beta^4 +18\beta^3 -18\beta^2+7\beta-1}{6\beta^3-23\beta^2 +26\beta-8}\Big).
\label{V.15}
\end{eqnarray}
then the field equations \eqref{a.2}-\eqref{a.4} for \eqref{V.13} are satisfied with a constant dilaton.

\subsubsection{ {Dualizing with respect to the coordinate $z$}}
\label{sec.5.1.3}

In what follows we shall dualize the G\"{o}del metrics on the coordinate $z$.
To this end, we choose the coupling matrices in the following form
{\small \begin{eqnarray}\label{VI.16}
F^{^{(1)}}_{a j}=\left( \begin{array}{ccc}
                    0~& - 2\mathbb{N} r ~&0
                      \end{array} \right),~F^{^{(2)}}_{i b}=\left( \begin{array}{c}
                    0  \\
                     2\mathbb{N} r\\
                    0
                      \end{array} \right),~F_{ij} = \left( \begin{array}{ccc}
                    \frac{l^2}{r^2}  & 0 & 0\\
                       0 & (\beta-1)l^2 r^2 & -r(l^2-2\mathbb{E})\\
                       0 & -r(l^2+2\mathbb{E}) & -l^2
                      \end{array} \right),
\end{eqnarray}}
and $E_{0\;ab}=l^2$. Then, by utilizing \eqref{c.5} one obtains
the background of the model in the same forms as \eqref{b.10} and \eqref{b.17}.
Analogously, only the $\beta =1$ case of this background with conditions $\mathbb{E}^2=l^4/4 $ and $\mathbb{N} =0$
can satisfy the vanishing of the one-loop beta-function equations in such a way that the dilaton field is
$\Phi = \phi^{^{(0)}} = f z+b$ in which $f^2=-1+ 2 l^2 \Lambda $.

The corresponding dual background can be cast in the form
\begin{eqnarray}
{d\tilde {s}}^2 &=& \frac{l^2 dr^2 }{r^2}-l^2 d t^2 + \frac{1}{l^2} d\tilde{x}^2 + \big((\beta-1) l^2 +\frac{4\mathbb{N}^2}{l^2}\big) r^2 d \varphi^2 -
\frac{4 \mathbb{N}}{l^2} r  d\tilde{x} d \varphi -2r l^2 dt d \varphi,\nonumber\\
{\tilde H} &=&\mathbb{E} dr \wedge d \varphi \wedge dt,\label{V.17}
\end{eqnarray}
where $\tilde x$ is the dualized coordinate of the dual manifold.
By taking into account the conditions $\mathbb{E}=l^2/2$ and  $\mathbb{N}=0$ one can show that
under the transformation $\tilde{x} \rightarrow l^2 z$, the background \eqref{V.17}
turns into the same forms as \eqref{b.10} and \eqref{b.12}. In this situation we are faced with self-duality again.
Imposing these conditions on \eqref{V.17} and  putting $\beta =1$ one shows that the dual
background is conformally invariant up to the one-loop order in a way that the dual dilaton is found by using equation \eqref{c.16.3} to be
${\tilde \Phi} =  f z+b-\log l^2$.

In addition, we check that the background \eqref{V.17} along with a constant dilaton
satisfy the vanishing of the two-loop beta-function equations if we have
\begin{eqnarray}
\mathbb{E}^2 = \frac{\beta^2 l^4}{\beta+1},~~~~~~~\mathbb{N}^2=\frac{l^4}{4},~~~~~~~\alpha' =- \frac{(\beta+1) l^2}{3\beta },
~~~~~~~{\tilde \Lambda}  = \frac{5-6\beta}{6l^2(\beta+1)}.\label{V.18}
\end{eqnarray}
Also, there is an additional solution with the dilaton field ${\tilde \Phi}={\tilde f}  l^2 z +b$ provided that
\begin{eqnarray}
\mathbb{\mathbb{E}}^2 = \frac{l^4}{4} (4 \beta -3),~~~~\mathbb{N}=0,~~~~~\alpha' =\frac{2 \beta l^2}{5-6 \beta},
~~~~~{\tilde f}^2 =- \frac{2}{\beta}+2\big({\tilde \Lambda} l^2 +1\big).\label{V.18.1}
\end{eqnarray}

\subsection{Abelian T-duality with two-dimensional Abelian Lie group}

\subsubsection{ {Dualizing with respect to both coordinates $(t, \varphi)$}}

In this subsection we shall perform the dualizing on both the coordinates $t$ and $\varphi$.
That is, the Lie group $G$ of the target manifold ${\cal M} \approx O \times  G$ is parameterized by
$g=e^{t T_{_1}} e^{\varphi  T_{_2}}$, wherein $(T_{_1} ,  T_{_2})$ are the basis of the Abelian Lie algebra $\cal G$ of $G$.
So the coordinates of the orbit $O$ are, in this case, represented by  $(r, z)$. In this respect, one may choose
the background matrices in the following forms
\begin{eqnarray}
E_{0\;ab}&=&\left( \begin{array}{cc}
                    -l^2 ~& -r(2\mathbb{E}+l^2)\\
                    r(2\mathbb{E}-l^2) ~  &(\beta -1)l^2 r^2
                      \end{array} \right),~~~~~~~~~
F^{^{(1)}}_{a j}=\left( \begin{array}{cc}
                    0 ~& 0\\
                    0 ~& 2\mathbb{N}r
                      \end{array} \right),\nonumber\\
F^{^{(2)}}_{i b}&=&\left( \begin{array}{cc}
                    0 ~& 0\\
                    0 ~& -2\mathbb{N}r
                      \end{array} \right),~~~~~~~~~~~~~~~~~~~~~~~~~~~~~~~~
F_{ij}=\left( \begin{array}{cc}
                    \frac{l^2}{r^2}~ & 0\\
                    0 ~& l^2
                      \end{array} \right).\label{V.19}
\end{eqnarray}
Hence, the background of the original $\sigma$-model implies the same metric \eqref{b.10} and field strength \eqref{b.17}.
From Sec. \ref{Sec.III} we know that the $\beta=1$  case of metric \eqref{b.10} along with $\mathbb{E}^2=l^4/4$ and $\mathbb{N}=0$
can be considered as a solution for the field equations \eqref{a.2}-\eqref{a.4} up to zeroth order in $\alpha'$.
In this situation,  the matrix $E_{0\;ab}$ will no longer be invertible, hence
the model can't be conformally invariant at one-loop order.

The corresponding  elements  to the dual model can be obtained by making use of relations  \eqref{V.19}
and \eqref{c.12}-\eqref{c.15}.
They are then read
\begin{eqnarray}
{d\tilde {s}}^2 &=& \frac{l^2 dr^2 }{r^2}+\frac{1}{4\mathbb{E}^2-\beta l^4} \Big[(\beta-1) l^2 d {\tilde x}_{_{1}}^2-\frac{l^2}{r^2} d  {\tilde x}_{_{2}}^2
+(4\mathbb{E}^2-4\mathbb{N}^2-\beta l^4) l^2 d z^2 \nonumber\\
&&~~~~~~~~~~~~~~~~~~~~~~~~~~~~~~~~~~~+\frac{2l^2}{r} d {\tilde x}_{_{1}} d {\tilde x}_{_{2}} +4\mathbb{N}l^2 (d {\tilde x}_{_{1}} dz - \frac{1}{r} d {\tilde x}_{_{2}} dz )\Big],\nonumber\\
{\tilde H} &=& -\frac{\mathbb{E}}{r^2(4\mathbb{E}^2-\beta l^4)} d {\tilde x}_{_{1}} \wedge d {\tilde x}_{_{2}} \wedge dr,\label{V.20}
\end{eqnarray}
where $({\tilde x}_{_{1}} , {\tilde x}_{_{2}})$ are the dualized coordinates of the dual manifold.
We have checked that there is no dilaton field to support the conformality of the dual background \eqref{V.20} at one-loop level.
But, one can check that the dual background along with a constant dilaton field
is conformally invariant up to the first order in $\alpha'$ provided that
\begin{eqnarray}
\mathbb{E}^2&=&\frac{l^4}{4},~~~~~~~\mathbb{N}^2=\frac{(1-\beta)(\beta^2+2\beta-3)l^4}{3\beta^2+7\beta-6},~~~~~~~
\alpha' = \frac{2\beta(3\beta -2)l^2}{2\beta^2+\beta-4},\nonumber\\
{\tilde \Lambda} &=& \frac{2\beta^4-6\beta^2 +3\beta+1}{\beta(3\beta-2) (2\beta^2+\beta-4)l^2}.\label{V.21}
\end{eqnarray}
Notice that this solution is valid for every $\beta \in (0 , 2/3)$.
It should be noted that by taking into consideration $\mathbb{N}=0$  and also condition $\beta \neq 1$ for \eqref{V.20}
we see that under the transformation
\begin{eqnarray}
{\tilde x}_{_{1}} = \sqrt{\frac{\beta l^4- 4\mathbb{E}^2 }{\beta-1}}~ t',~~~~~~{\tilde x}_{_{2}} = \varphi',~~~~~~
r=\frac{1}{r' \sqrt{(\beta -1)({\beta l^4  - 4\mathbb{E}^2})}},~~~~~~z=z',\label{V.22}
\end{eqnarray}
the dual background turns into the same metric \eqref{b.10} and field strength \eqref{b.12},
which this is nothing but the self-duality of the G\"{o}del metrics.

\subsubsection{ {Dualizing with respect to both coordinates $(t, z)$}}

When the dualizing is performed on the coordinates $(t, z)$,
it is more appropriate to choose the spectator-dependent matrices as follows:
\begin{eqnarray}
E_{0\;ab}&=&\left( \begin{array}{cc}
                    -l^2 ~& 0\\
                    0   ~& l^2
                      \end{array} \right),~~~~~~~~~~~~~~~~~
F^{^{(1)}}_{a j}=\left( \begin{array}{cc}
                    0 ~& -r(2\mathbb{E}+l^2)\\
                    0 ~& - 2\mathbb{N}r
                      \end{array} \right),\nonumber\\
F^{^{(2)}}_{i b}&=&\left( \begin{array}{cc}
                    0 ~& 0\\
                    r(2\mathbb{E}-l^2) ~& 2\mathbb{N}r
                      \end{array} \right),~~~~~~
F_{ij}=\left( \begin{array}{cc}
                    \frac{l^2}{r^2} ~& 0\\
                    0 ~& (\beta -1)r^2l^2
                      \end{array} \right).\label{V.23}
\end{eqnarray}
With regard to this choice, we arrive at familiar results, i.e., equations \eqref{b.10} and  \eqref{b.17}.
According to Sec. \ref{Sec.III}, the case of $\beta =1$ of this background with conditions $\mathbb{E}^2=l^4/4 $ and $\mathbb{N} =0$
is a solution for the one-loop beta-function equations in such a way that the dilaton field is obtained to be
$\Phi = \phi^{^{(0)}} = f z+b$ in which $f^2=-1+ 2 l^2 \Lambda $.

Now, one applies formulae \eqref{c.12}-\eqref{c.15} and \eqref{V.23} to obtain the dual background in the following form
\begin{eqnarray}
{d\tilde {s}}^2 &=& \frac{l^2 dr^2 }{r^2} + \frac{1}{l^2} \Big[-d {\tilde x}_{_{1}}^2+ d  {\tilde x}_{_{2}}^2
-(4\mathbb{E}^2-4\mathbb{N}^2-\beta l^4) r^2 d \varphi^2 +4r (\mathbb{E} d {\tilde x}_{_{1}} d\varphi - \mathbb{N} d {\tilde x}_{_{2}} d\varphi)\Big], \nonumber\\
{\tilde H} &=& -\frac{1}{2} d {\tilde x}_{_{1}} \wedge d r \wedge d\varphi.\label{V.24}
\end{eqnarray}
Analogously, $({\tilde x}_{_{1}} , {\tilde x}_{_{2}})$ are the dualized coordinates.
One can check that only a particular case of the dual background is conformally invariant up to the one-loop order, giving
\begin{eqnarray}\label{V.24.1}
{d\tilde {s}}^2 &=& \frac{l^2 dr^2 }{r^2} - {l^2} d t^2+ l^2 d  z^2
 -2 r l^2 d t d\varphi, \nonumber\\
{\tilde H} &=& -\frac{l^2}{2} d t \wedge d r \wedge d\varphi,\nonumber\\
{\tilde \Phi} &=&  f z+b-\log l^4,
\end{eqnarray}
where $f^2=-1+ 2 l^2 \Lambda $.  We have used the transformation ${\tilde x}_{_{1}} \rightarrow l^2 t,~{\tilde x}_{_{2}} \rightarrow l^2 z$, and
have put $\beta =1$, $\mathbb{E} = -l^2/2$ and $\mathbb{N}=0$; moreover,
the dual dilaton have been obtained  by using \eqref{c.16.3}.

In addition, a constant dilaton field guaranties the conformal invariance of the dual background \eqref{V.24}
up to the first order in $\alpha'$
if the constants $\mathbb{E}, \mathbb{N}, l, \beta, {\tilde \Lambda}$ and $\alpha'$ satisfy the following relations
\begin{eqnarray}
\mathbb{E}^2&=&2{l^4},~~~~~~~~~~\mathbb{N}^2=\frac{l^4}{24} (\Theta -7\beta+50),~~~~~~~~~~
\alpha' = \frac{24 \beta l^2}{7\beta -\Theta -20},\nonumber\\
{\tilde \Lambda} &=& \frac{-59 \beta^2 +(236+5\Theta) \beta +20\Theta-110}{24 \beta l^2 (\Theta -7\beta +20)},\label{V.25}
\end{eqnarray}
where $\Theta = \sqrt{\beta^2 -16 \beta +100}$.

\subsubsection{ {Dualizing with respect to both coordinates $(\varphi, z)$}}

It is also possible to perform the dualizing on both the coordinates $\varphi$ and $z$.
So the coordinates of orbit $O$ are chosen to be $(t, r)$.
In this way, both the G\"{o}del metrics \eqref{b.10} and field strength \eqref{b.17} may be yielded from
the original $\sigma$-model if one considers
\begin{eqnarray}
E_{0\;ab}&=&\left( \begin{array}{cc}
                    l^2~ & - 2\mathbb{N}r\\
                     2\mathbb{N}r ~  & (\beta -1)r^2l^2
                      \end{array} \right),~~~~~~~
F^{^{(1)}}_{a j}=\left( \begin{array}{cc}
                    0 ~& 0\\
                    0 ~& r(2\mathbb{E}-l^2)
                      \end{array} \right),\nonumber\\
F^{^{(2)}}_{i b}&=&\left( \begin{array}{cc}
                    0 ~& 0\\
                    0 ~& -r(2\mathbb{E}+l^2)
                      \end{array} \right),~~~~~~~~~~~
F_{ij}=\left( \begin{array}{cc}
                    \frac{l^2}{r^2}~ & 0\\
                    0 ~& -l^2
                      \end{array} \right).\label{V.26}
\end{eqnarray}
The one-loop conformal invariance conditions of the G\"{o}del metrics require that we have $\beta=1, \mathbb{E}^2 =l^4/4$ and $ \mathbb{N}=0$. In this situation
the $E_0$ is singular. Thus, when the dualizing is implemented by the shift of directions $(\varphi, z)$,
the pair of the mutually dual models can't be conformally invariant at one-loop level.

Similar to the preceding cases one obtains the corresponding dual $\sigma$-model. If the dualized coordinates are
considered to be $({\tilde x}_{_{1}} , {\tilde x}_{_{2}})$, then we have
\begin{eqnarray}
{d\tilde {s}}^2 &=& \frac{l^2 dr^2 }{r^2}+\frac{1}{4\mathbb{N}^2+(\beta-1) l^4} \Big[(\beta-1) l^2 d {\tilde x}_{_{1}}^2+\frac{l^2}{r^2} d  {\tilde x}_{_{2}}^2
+(4\mathbb{E}^2-4\mathbb{N}^2-\beta l^4) l^2 d t^2 \nonumber\\
&&~~~~~~~~~~~~~~~~~~~~~~~~~~~~~~~~~~~~~~~~~~~~~~~~~~~~~~~~~-4l^2 (\mathbb{N}d {\tilde x}_{_{1}} dt -\frac{\mathbb{E}}{r} d {\tilde x}_{_{2}} dt)\Big],\nonumber\\
{\tilde H} &=& \frac{1}{r^2\big[4\mathbb{N}^2+(\beta-1) l^4\big]} \Big(\frac{l^4}{2} dr \wedge  d {\tilde x}_{_{2}} \wedge dt -\mathbb{N}
dr \wedge d {\tilde x}_{_{1}}\wedge  d {\tilde x}_{_{2}}\Big).\label{V.27}
\end{eqnarray}
It's worth mentioning that under the transformation
${\tilde x}_{_{1}} = l^2 z',~{\tilde x}_{_{2}} =\varphi',~r=1/\big((\beta -1)l^2 r'\big),~t=-t'$,
the background \eqref{V.27} can be reduced to the G\"{o}del metrics \eqref{b.10} and field strength \eqref{b.12}; of course when we put $\mathbb{N}=0$ and $\mathbb{E}=l^2/2$.

\subsection{Abelian T-duality with three-dimensional Abelian Lie group}

There is a possibility that we can perform the dualizing on the coordinates
$x^\mu=(z, t, \varphi)$. In fact, these are the coordinates of the Abelian Lie group of the target space.
Choosing the appropriate spectator-dependent matrices in the forms
\begin{eqnarray}\label{V.28}
E_{0\;ab} =\left( \begin{array}{ccc}
                    l^2  & 0 & -2\mathbb{N} r\\
                       0 & -l^2 & -(l^2 +2\mathbb{E})r\\
                       2\mathbb{N} r & -(l^2-2\mathbb{E})r & (\beta-1)l^2 r^2
                      \end{array} \right),~~F^{^{(1)}}_{a j}=0,~~F^{^{(2)}}_{i b}=0,~~F_{ij} =\frac{l^2}{r^2},
\end{eqnarray}
one gets the coupling matrices of the original $\sigma$-model. Then, by making use of \eqref{c.5}
the metric and field strength of the model are obtained to be of the same forms as \eqref{b.10} and \eqref{b.17}, respectively.
Because of the invertibility of the matrix $E_{0}$, we can't accept the values $\beta=1$, $\mathbb{E}^2=l^4/4$ and $\mathbb{N}=0$
for $E_{0}$.
Accordingly, the original $\sigma$-model doesn't satisfy the one-loop beta-function equations.

The dual background can be cast in the form
\begin{eqnarray}
{d\tilde {s}}^2 &=& \frac{l^2 }{r^2}dr^2 -\frac{1}{4(\mathbb{E}^2-\mathbb{N}^2) -\beta l^4}~\Big[(\beta l^2 -\frac{4\mathbb{E}^2}{l^2}) d {\tilde x}_{_{1}}^2-
\big((\beta-1) l^2 +\frac{4\mathbb{N}^2}{l^2}) d {\tilde x}_{_{2}}^2\nonumber\\
&&~~+\frac{l^2}{r^2} d {\tilde x}_{_{3}}^2 + \frac{8\mathbb{E}\mathbb{N}}{l^2} d {\tilde x}_{_{1}} d {\tilde x}_{_{2}}
- \frac{2l^2}{r} d {\tilde x}_{_{2}} d {\tilde x}_{_{3}}\Big],\nonumber\\
{\tilde H} &=& -\frac{1}{r^2 \Big(4(\mathbb{E}^2-\mathbb{N}^2) -\beta l^4\Big)} \Big[\mathbb{E} dr \wedge d {\tilde x}_{_{2}} \wedge d {\tilde x}_{_{3}}
+\mathbb{N} dr \wedge d {\tilde x}_{_{3}} \wedge d {\tilde x}_{_{1}}\Big],\label{V.29}
\end{eqnarray}
where $({\tilde x}_{_{1}}, {\tilde x}_{_{2}}, {\tilde x}_{_{3}})$ are the dualized coordinates.
One can easily check that the dual background is not also conformally invariant up to the one-loop order.
It is quite interesting to comment on the $\beta=1/2$ case of the background \eqref{V.29}. For this case, one can easily show that
the field equations \eqref{a.2}-\eqref{a.4} are satisfied  with a constant dilaton provided that
\begin{eqnarray}
\mathbb{E} = 0,~~~~~~~~\mathbb{N}^2 =\frac{l^4}{8},~~~~~~~~~\alpha' = 4 l^2,~~~~~~~~~
{\tilde \Lambda} = \frac{3}{8 l^2 }.\label{V.31}
\end{eqnarray}
Putting $\mathbb{N}=0$ in both the metric and corresponding  $B$-field of the background \eqref{V.29} and then utilizing the transformation
\begin{eqnarray}
{\tilde x}_{_{1}} = {l^2} z',~~~~{\tilde x}_{_{2}} = \sqrt{\frac{\beta l^4- 4\mathbb{E}^2}{\beta-1}} ~t',~~~~~
{\tilde x}_{_{3}} = {\sqrt{(\beta-1)(\beta l^4- 4\mathbb{E}^2)}} ~\varphi',~~~~r=\frac{1}{r'},\label{V.32}
\end{eqnarray}
the dual background turns into the same metric \eqref{b.10} and field strength \eqref{b.12}.

In summary, we obtained seven Abelian duals for the G\"{o}del spacetimes
and showed that they were self-dual when $\mathbb{N}$ went to zero,
and in some instances, one must consider $\mathbb{E}=l^2/2$.
The results of this section are clearly summarized in Table 1. There, we have shown the backgrounds solving the one-loop equations, and
solve also the two-loop equations without $\alpha'$-corrections.

 \begin{table}
			\caption{ Abelian T-duality of the G\"{o}del string cosmologies without corrections}
			\centering
			\begin{tabular}{| p{4em} | p{19em} | p{6.75em} | p{7.25em}|}
				\hline
				\multicolumn{2}{|c|}{\scriptsize  Dualizing with respect to the $\varphi$ coordinate}  & \scriptsize  Conditions for the one-loop solution & \scriptsize Conditions for the two-loop solution\\
				\hline

				\multirow{3}{4em}{\scriptsize Original background}&{\scriptsize $ds^2=l^2\big(-d{t}^2 + \frac{dr^2}{r^2}+(\beta-1)r^2~ d{\varphi}^2-2r~ dt d\varphi +d{z}^2\big)$}&  \multirow{3}{6.75em}{-----}   & \multirow{3}{7.25em}{\scriptsize $\beta\neq1$ with the conditions given in Eqs. \eqref{b.18} and \eqref{b.19} } \\
				
				&{\scriptsize $B=2r(\mathbb{E} d{\varphi} \wedge dt + \mathbb{N}d{\varphi} \wedge dz)$} && \\
				
				&{\scriptsize $ \Phi=b$}&& \\
				
				\hline

				\multirow{4}{4em}{\scriptsize Dual background}&\multirow{4}{19em}{\scriptsize ${d\tilde {s}}^2=\frac{l^2 dr^2 }{r^2}+ \frac{1}{(\beta -1)}
					\Big[(\frac{4 \mathbb{E}^2}{l^2}-\beta  l^2) dt^2 +\frac{d\tilde{x}^2}{ l^2 r^2} +\big(\frac{4 \mathbb{N}^2}{ l^2}+l^2 (\beta -1)\big)dz^2+\frac{8\mathbb{E} \mathbb{N}}{l^2}  dt dz+\frac{4\mathbb{E}}{l^2 r} d\tilde{x} dt+\frac{4\mathbb{N}}{l^2 r} d\tilde{x} dz\Big]$}&\multirow{7}{6.75em}{-----}   &{ \scriptsize  $\mathbb{N}=0,\alpha'=\frac{6l^2\beta}{2\beta-5},$} \\
				
				&&&{ \scriptsize $\mathbb{E}^2=\frac{l^4(4\beta-1)}{12},$} \\
				
				& & &{ \scriptsize $\tilde f^2=\frac{2(1+\beta(3l^2\tilde\Lambda-1))}{3\beta}$} \\
				\cline{4-4}
				&\scriptsize ${\tilde B} = \frac{1}{\beta-1} (\frac{1}{r}dt \wedge  d\tilde{x}+2 \mathbb{N}  dt\wedge dz)$ & &\scriptsize $\mathbb{E}^2=\frac{l^4(2 \beta^2 -2 \beta +1)}{4},$  \\
				&\scriptsize $\tilde\Phi=\tilde fz+b$ & &\scriptsize $\mathbb{N}^2=\frac{l^4(\beta -1)^2}{2},$  \\
				& & &\scriptsize $\alpha' = \frac{2 l^2 \beta}{2 -3\beta },\tilde f=0,$  \\
				& & &\scriptsize $\tilde \Lambda = \frac{(\beta -1)^2}{4 l^2 \beta (2- 3{\beta})}$  \\
				\hline
			\end{tabular}

				\begin{tabular}{| p{4em} | p{19em} | p{6.75em} | p{7.25em}|}
					\hline
					\multicolumn{2}{|c|}{\scriptsize  Dualizing with respect to the $t$ coordinate}  & \scriptsize  Conditions for the one-loop solution & \scriptsize Conditions for the two-loop solution\\
					\hline

					\multirow{3}{4em}{\scriptsize Original background}&{\scriptsize $ds^2=l^2\big(-d{t}^2 + \frac{dr^2}{r^2}+(\beta-1)r^2~ d{\varphi}^2-2r~ dt d\varphi +d{z}^2\big)$}& \scriptsize $\mathbb{E}^2=\frac{l^4}{4},~\mathbb{N}=0,$ & \scriptsize $\mathbb{E}^2=\frac{l^4(4\beta-3)}{4},$\\
					
					&{\scriptsize $B=2r(\mathbb{E} d{\varphi} \wedge dt + \mathbb{N}d{\varphi} \wedge dz)$} &\scriptsize $\beta=1,$ &\scriptsize $\alpha'=\frac{2l^2\beta}{5-6\beta},~\mathbb{N}=0,$ \\
					
					&{\scriptsize $ \Phi=fz+b$}&\scriptsize $f^2=2l^2\Lambda-1$ &\scriptsize $f^2=2(1+l^2\Lambda-\frac{1}{\beta}$) \\
					\hline
					
					\multirow{6}{4em}{\scriptsize Dual background}&\multirow{3}{19em}{\scriptsize ${d\tilde {s}}^2=\frac{l^2 dr^2 }{r^2}+l^2 d z^2 - \frac{1}{l^2} d\tilde{x}^2 + (\beta l^2 -\frac{4\mathbb{E}^2}{l^2}) r^2 d \varphi^2 +
						\frac{4 \mathbb{E}}{l^2} r  d\tilde{x} d \varphi$}& &{ \scriptsize  $\mathbb{N}=0,~\beta=1,$} \\
					
					&&\scriptsize $\beta=1,~\mathbb{N}=0,$ &{ \scriptsize $\mathbb{E}^2=\frac{l^4}{4},$} \\
					
					&\scriptsize ${\tilde B} = -r d\phi \wedge  d\tilde{x}+2 \mathbb{N} r d\phi\wedge dz$&\scriptsize $\mathbb{E}^2=\frac{l^4}{4},$ &{ \scriptsize $\tilde f^2=\frac{-\alpha'}{2l^2}+2l^2\tilde\Lambda-1$} \\
					\cline{4-4}
					& \scriptsize $\tilde\Phi=\tilde fz+b-\log l^2$ &\scriptsize $\tilde f^2=2l^2\tilde\Lambda-1$ &\scriptsize $\tilde f=0$~with the conditions given in \eqref{V.15}  \\

					\hline
				\end{tabular}
				\begin{tabular}{| p{4em} | p{19em} | p{6.75em} | p{7.25em}|}
					\hline

					\multicolumn{2}{|c|}{\scriptsize  Dualizing with respect to the $z$ coordinate}  & \scriptsize  Conditions for the one-loop solution & \scriptsize Conditions for the two-loop solution\\
					\hline

					\multirow{3}{4em}{\scriptsize Original background}&{\scriptsize $ds^2=l^2\big(-d{t}^2 + \frac{dr^2}{r^2}+(\beta-1)r^2~ d{\varphi}^2-2r~ dt d\varphi +d{z}^2\big)$}& \scriptsize $\mathbb{E}^2=\frac{l^4}{4},\mathbb{N}=0,$ & \scriptsize $\mathbb{E}^2=\frac{l^4(4\beta-3)}{4},$\\
					
					&{\scriptsize $B=2r(\mathbb{E} d{\varphi} \wedge dt + \mathbb{N}d{\varphi} \wedge dz)$} &\scriptsize $\beta=1,$ &\scriptsize $\alpha'=\frac{2l^2\beta}{5-6\beta},~\mathbb{N}=0,$ \\
					
					&{\scriptsize $ \Phi=fz+b$}&\scriptsize $f^2=2l^2\Lambda-1$ &\scriptsize $f^2=2(1+l^2\Lambda-\frac{1}{\beta}$) \\
					\hline
					
					\multirow{6}{4em}{\scriptsize Dual background}&\multirow{4}{19em}{\scriptsize ${d\tilde {s}}^2=\frac{l^2 dr^2 }{r^2}-l^2 d t^2 + \frac{1}{l^2} d\tilde{x}^2 + \big((\beta-1) l^2 +\frac{4\mathbb{N}^2}{l^2}\big) r^2 d \varphi^2 -
						\frac{4 \mathbb{N}}{l^2} r  d\tilde{x} d \varphi -2r l^2 dt d \varphi$}& &{ \scriptsize  $\mathbb{N}=0,~\alpha' =\frac{2 \beta l^2}{5-6 \beta},$} \\
					
					&&\scriptsize $\beta=1,~\mathbb{N}=0,$ &{ \scriptsize $\mathbb{E}^2 = \frac{l^4}{4} (4 \beta -3)$,} \\
					
					& &\scriptsize $\mathbb{E}^2=\frac{l^4}{4},$ &{ \scriptsize ${\tilde f}^2 =- \frac{2}{\beta}+2\big({\tilde \Lambda} l^2 +1\big)$} \\
					\cline{4-4}
					&\scriptsize ${\tilde B} = 2 \mathbb{E} r d\varphi\wedge dt$ &\scriptsize $\tilde f^2=2l^2\tilde\Lambda-1$ &\scriptsize $\mathbb{N}^2=\frac{l^4}{4},~\tilde f=0,$  \\
					
					&\scriptsize $\tilde\Phi=\tilde fz+b-\log l^2$& &\scriptsize $\mathbb{E}^2 = \frac{\beta^2 l^4}{\beta+1},$ \\
					&& &\scriptsize $\alpha' =- \frac{(\beta+1) l^2}{3\beta },$ \\
					&& &\scriptsize ${\tilde \Lambda}  = \frac{5-6\beta}{6l^2(\beta+1)}$ \\
					
					\hline 		  		
				\end{tabular}
	\end{table}
\begin{table}
	\caption*{ Table 1: Continued}
				\begin{tabular}{| p{4em} | p{19em} | p{6.75em} | p{7.25em}|}
					\hline
					\multicolumn{2}{|c|}{\scriptsize  Dualizing with respect to the $(t,\varphi)$ coordinates}  & \scriptsize  Conditions for the one-loop solution & \scriptsize Conditions for the two-loop solution\\
					\hline

					\multirow{3}{4em}{\scriptsize Original background}&{\scriptsize $ds^2=l^2\big(-d{t}^2 + \frac{dr^2}{r^2}+(\beta-1)r^2~ d{\varphi}^2-2r~ dt d\varphi +d{z}^2\big)$}&  \multirow{3}{6.75em}{-----}   & \multirow{3}{7.25em}{\scriptsize $\beta\neq1$ with the conditions given in Eqs. \eqref{b.18} and \eqref{b.19} } \\
					
					&{\scriptsize $B=2r(\mathbb{E} d{\varphi} \wedge dt + \mathbb{N}d{\varphi} \wedge dz)$} && \\
					
					&{\scriptsize $ \Phi=b$}&& \\
					
					\hline

					\multirow{4}{4em}{\scriptsize Dual background}&\multirow{4}{19em}{\scriptsize ${d\tilde {s}}^2=\frac{l^2 dr^2 }{r^2}+\frac{1}{4\mathbb{E}^2-\beta l^4} \Big[(\beta-1) l^2 d {\tilde x}_{_{1}}^2-\frac{l^2}{r^2} d  {\tilde x}_{_{2}}^2+(4\mathbb{E}^2-4\mathbb{N}^2-\beta l^4) l^2 d z^2+\frac{2l^2}{r} d {\tilde x}_{_{1}} d {\tilde x}_{_{2}} +4\mathbb{N}l^2 (d {\tilde x}_{_{1}} dz - \frac{1}{r} d {\tilde x}_{_{2}} dz )\Big]$}&\multirow{6}{6.75em}{-----}   &{ \scriptsize  $\mathbb{N}=0,~\alpha'=\frac{2\beta l^2}{5-6\beta},$} \\
					
					&&&{ \scriptsize $\mathbb{E}^2=\frac{l^4(4\beta-3)}{4},$} \\
					
					& & &{ \scriptsize $\tilde f^2=\frac{2(\beta-1+l^2\tilde\Lambda\beta)}{\beta}$} \\
					\cline{4-4}
					&\scriptsize ${\tilde B} = \frac{1}{r^2(4\mathbb{E}^2-\beta l^4)} \big(2\mathbb{E}r d{\tilde x}_{_{1}} \wedge d{\tilde x}_{_{2}}+4\mathbb{E}\mathbb{N}r^2 d{\tilde x}_{_{1}} \wedge dz\big)$ & &\scriptsize $\tilde f=0$ with the conditions given in Eq. \eqref{V.21}  \\
					&\scriptsize $\tilde\Phi=\tilde fz+b$ & & \\
					\hline
				\end{tabular}
		
				\begin{tabular}{| p{4em} | p{19em} | p{6.75em} | p{7.25em}|}
					\hline
					\multicolumn{2}{|c|}{\scriptsize  Dualizing with respect to the $(t,z)$ coordinates}  & \scriptsize  Conditions for the one-loop solution & \scriptsize Conditions for the two-loop solution\\
					\hline

					\multirow{5}{4em}{\scriptsize Original background}&& & \scriptsize $\mathbb{E}^2=\frac{l^4(4\beta-3)}{4},$\\
					
					&{\scriptsize $ds^2=l^2\big(-d{t}^2 + \frac{dr^2}{r^2}+(\beta-1)r^2~ d{\varphi}^2-2r~ dt d\varphi +d{z}^2\big)$} & \scriptsize $\mathbb{E}^2=\frac{l^4}{4},~\mathbb{N}=0,$ &\scriptsize $\alpha'=\frac{2l^2\beta}{5-6\beta},~\mathbb{N}=0,$ \\
					
					&{\scriptsize $B=2r(\mathbb{E} d{\varphi} \wedge dt + \mathbb{N}d{\varphi} \wedge dz)$}&\scriptsize $\beta=1,$&\scriptsize $f^2=2(1+l^2\Lambda-\frac{1}{\beta}$) \\
					\cline{4-4}
					&{\scriptsize $ \Phi=fz+b$}&\scriptsize $f^2=2l^2\Lambda-1$ &\scriptsize $f=0 $ with the conditions given in Eqs. \eqref{b.18} and \eqref{b.19} \\
					\hline
					
					\multirow{6}{4em}{\scriptsize Dual background}&\multirow{4}{19em}{\scriptsize ${d\tilde {s}}^2=\frac{l^2 dr^2 }{r^2} + \frac{1}{l^2} \Big[-d {\tilde x}_{_{1}}^2+ d  {\tilde x}_{_{2}}^2
						-(4\mathbb{E}^2-4\mathbb{N}^2-\beta l^4) r^2 d \varphi^2 +4r (\mathbb{E} d {\tilde x}_{_{1}} d\varphi - \mathbb{N} d {\tilde x}_{_{2}} d\varphi)\Big]$}& &{ \scriptsize  $\mathbb{N}=0,~\alpha' =\frac{6 \beta l^2}{2 \beta-5},$} \\
					
					&&\scriptsize $\beta=1,~\mathbb{N}=0,$ &{ \scriptsize $\mathbb{E}^2 = \frac{l^4}{12} (4 \beta -1)$,} \\
					
					& &\scriptsize $\mathbb{E}=-\frac{l^2}{2},$ &\scriptsize ${\tilde f}^2 = \frac{2\big(1+\beta(3l^2\tilde\Lambda-1)\big)}{3\beta l^4}$ \\
					\cline{4-4}
					&\scriptsize ${\tilde B} = r d {\tilde x}_{_{1}}\wedge d\varphi$ &\scriptsize $\tilde f^2=2l^2\tilde\Lambda-1$ &\scriptsize $\tilde f=0$ with the conditions given in Eq. \eqref{V.25}  \\
					
					&\scriptsize $\tilde\Phi=\tilde fz+b$& & \\
					
					\hline 		  		
				\end{tabular}
				\label{tab:LPer}
				\begin{tabular}{| p{4em} | p{19em} | p{6.75em} | p{7.25em}|}
					\hline
					\multicolumn{2}{|c|}{\scriptsize  Dualizing with respect to the $(\varphi,z)$ coordinates}  & \scriptsize  Conditions for the one-loop solution & \scriptsize Conditions for the two-loop solution\\
					\hline

					\multirow{3}{4em}{\scriptsize Original background}&{\scriptsize $ds^2=l^2\big(-d{t}^2 + \frac{dr^2}{r^2}+(\beta-1)r^2~ d{\varphi}^2-2r~ dt d\varphi +d{z}^2\big)$}&  \multirow{3}{6.75em}{-----}   & \multirow{3}{7.25em}{\scriptsize $\beta\neq1$  with the conditions given in Eqs. \eqref{b.18} and \eqref{b.19} } \\
					
					&{\scriptsize $B=2r(\mathbb{E} d{\varphi} \wedge dt + \mathbb{N}d{\varphi} \wedge dz)$} && \\
					
					&{\scriptsize $ \Phi=b$}&& \\
					
					\hline

					\multirow{4}{4em}{\scriptsize Dual background}&\multirow{2}{19em}{\scriptsize ${d\tilde {s}}^2=\frac{l^2 dr^2 }{r^2}+\frac{1}{4\mathbb{N}^2+(\beta-1) l^4} \Big[(\beta-1) l^2 d {\tilde x}_{_{1}}^2+\frac{l^2}{r^2} d  {\tilde x}_{_{2}}^2
						+(4\mathbb{E}^2-4\mathbb{N}^2-\beta l^4) l^2 d t^2 -4l^2 (\mathbb{N}d {\tilde x}_{_{1}} dt -\frac{\mathbb{E}}{r} d {\tilde x}_{_{2}} dt)\Big]$}&\multirow{4}{6.75em}{-----}   &{ \scriptsize  $\mathbb{N}=0,~\alpha'=\frac{2\beta l^2}{2\beta-3},$} \\
					
					&&&{ \scriptsize $\mathbb{E}^2=\frac{l^4}{4},$} \\
					
					&\scriptsize ${\tilde B} = \frac{1}{4\mathbb{N}^2+(\beta-1) l^4} (4\mathbb{E}\mathbb{N} d{\tilde x}_{_{1}} \wedge dt+\frac{2\mathbb{N}}{r}d{\tilde x}_{_{1}} \wedge d{\tilde x}_{_{2}}-\frac{l^4}{r}d{\tilde x}_{_{2}}\wedge dt)$  & &{ \scriptsize $\tilde \Lambda=\frac{(\beta-1)^2}{\beta l^2(2\beta-3)}$} \\
					
					&\scriptsize $\tilde\Phi=b$& & \\
					
					\hline
				\end{tabular}
			\end{table}

		 \begin{table}
		 	\caption*{ Table 1: Continued}
					\begin{tabular}{| p{4em} | p{19em} | p{6.75em} | p{7.25em}|}
						\hline
						\multicolumn{2}{|c|}{\scriptsize  Dualizing with respect to the $(t,\varphi,z)$ coordinates}  & \scriptsize  Conditions for the one-loop solution & \scriptsize Conditions for the two-loop solution\\
						\hline

						\multirow{3}{4em}{\scriptsize Original background}&{\scriptsize $ds^2=l^2\big(-d{t}^2 + \frac{dr^2}{r^2}+(\beta-1)r^2~ d{\varphi}^2-2r~ dt d\varphi +d{z}^2\big)$}&  \multirow{3}{6.75em}{-----}   & \multirow{3}{7.25em}{\scriptsize $\beta\neq1$ with the conditions given in Eqs. \eqref{b.18} and \eqref{b.19} } \\
						
						&{\scriptsize $B=2r(\mathbb{E} d{\varphi} \wedge dt + \mathbb{N}d{\varphi} \wedge dz)$} && \\
						
						&{\scriptsize $ \Phi=b$}&& \\
						
						\hline

						\multirow{6}{4em}{\scriptsize Dual background}&\multirow{3}{19em}{\scriptsize ${d\tilde {s}}^2=\frac{l^2 }{r^2}dr^2 -\frac{1}{\tilde\Delta}~\Big[(\beta l^2 -\frac{4\mathbb{E}^2}{l^2}) d {\tilde x}_{_{1}}^2-
							\big((\beta-1) l^2 +\frac{4\mathbb{N}^2}{l^2}) d {\tilde x}_{_{2}}^2+\frac{l^2}{r^2} d {\tilde x}_{_{3}}^2 + \frac{8\mathbb{E}\mathbb{N}}{l^2} d {\tilde x}_{_{1}} d {\tilde x}_{_{2}}
							- \frac{2l^2}{r} d {\tilde x}_{_{2}} d {\tilde x}_{_{3}}\Big]$}&\multirow{6}{6.75em}{-----}   &{ \scriptsize  $\beta=\frac{1}{2},~\tilde f=0,~\mathbb{E}=0,$} \\
						
						&&&{ \scriptsize $\mathbb{N}^2=\frac{l^4}{8},~\alpha'=4l^2,$} \\
						
						&  & &{ \scriptsize $\tilde \Lambda=\frac{3}{8l^2}$} \\
						\cline{4-4}
						
						&\scriptsize ${\tilde B} = \frac{1}{\tilde\Delta} \big(2\mathbb{N} d{\tilde x}_{_{1}} \wedge d{\tilde x}_{_{2}}-\frac{2\mathbb{N}}{r} d{\tilde x}_{_{1}} \wedge d{\tilde x}_{_{3}}+\frac{2\mathbb{E}}{r} d{\tilde x}_{_{2}} \wedge d{\tilde x}_{_{3}}\big)$& &\scriptsize  $\mathbb{N}=0,~\alpha'=\frac{2l^2\beta}{5-6\beta},$ \\
						&\scriptsize $\tilde\Phi=\tilde fz+b$& &\scriptsize  $\mathbb{E}^2=\frac{l^4(4\beta-3)}{4},$ \\
						&\scriptsize $\tilde\Delta=4(\mathbb{E}^2-\mathbb{N}^2) -\beta l^4$& &\scriptsize  $\tilde f^2=\frac{2(\beta-1+l^2\beta\tilde\Lambda)}{\beta l^4}$ \\
						\hline
					\end{tabular}
				\end{table}

Before closing this section, we note the fact
that our Abelian dual models were obtained by the PL T-duality transformations at the classical level.
The PL T-duality transformations are valid only at the one-loop level.
Our results showed that by applying the rules of Abelian
T-duality without further corrections, we were still able to obtain two-loop
solutions.
\section{\label{Sec.VI} Abelian T-duality of G\"{o}del string cosmologies up to $\alpha'$-corrections}

In this section, we study the Abelian T-duality of G\"{o}del string cosmologies up to $\alpha'$-corrections
when the dualizing is implemented by the shift of directions $z$ and $t$.
To this end, we use the T-duality rules at two-loop order derived by KM \cite{KM}.
Before proceeding further, let us review  the $\alpha'$-corrected T-duality
rules of KM.

\subsection{The $\alpha'$-corrections from T-duality rules at two-loop }

As mentioned in the introduction section, the authors of Ref. \cite{KM}
had obtained the two-loop $\sigma$-model corrections to the
T-duality map in string theory by using the effective action approach.
They had found the explicit form for the ${\cal O } {(\alpha')}$ modifications
of the lowest order duality transformations by focusing on backgrounds that
have a single Abelian isometry.
Recently, it has been shown that \cite{Wulff} using the $\alpha'$-corrected T-duality
rules of \cite{KM} one can obtain explicit $\alpha'$-corrections for Yang-Baxter deformed models.
In fact, it has been argued that (homogeneous) Yang-Baxter deformed string $\sigma$-models that are conformal at one
loop remain conformal at two loops.

Similar to  Ref. \cite{Wulff} we follow \cite{KM} and choose the reduced metric $g_{{\mu\nu}}$, antisymmetric field $b_{{\mu\nu}}$
and dilaton $\Phi$ of the $d$-dimensional spacetime, respectively, according to
\begin{eqnarray}
d s^2&=&G_{_{MN}} dX^{^{M}} dX^{^{N}} = g_{{\mu\nu}} dx^{\mu} dx^{\nu} + e^{2 \sigma}~ (d\underline{x} +V_{\mu} dx^{\mu})^2,\label{6.1}\\
B &=& \frac{1}{2} B_{_{MN}} dX^{^{M}} \wedge dX^{^{N}} = \frac{1}{2} b_{\mu\nu} dx^{\mu} \wedge dx^{\nu}
+\frac{1}{2} W_{\mu} V_{\nu}  dx^{\mu} \wedge dx^{\nu} + W_{\mu}   dx^{\mu} \wedge d\underline{x},\label{6.2}\\
\Phi &=& \hat{\phi}  +\frac{1}{2} \sigma,\label{6.3}
\end{eqnarray}
where the  field $V_{\mu}$ is the standard Kaluza-Klein
gauge field coupling to the momentum modes of the theory, and $W_{\mu}$ is the other gauge field,
which couples to the winding modes. Here we assume that  the isometry we
want to dualize is simply implemented by a shift of a coordinate, that we denote by $\underline{x}$.
We have also introduced the scalars ${\sigma}$ and $\hat{\phi} $.
In components, the relations to identify the fields of the dimensional reduction are
\begin{eqnarray}
g_{{\mu\nu}}&=& G_{{\mu\nu}} - \frac{G_{{\mu \underline{x}}} G_{{\nu \underline{x}}}}{G_{{\underline{x}  \underline{x}}}},
~~~~b_{{\mu\nu}}= B_{{\mu\nu}} + \frac{G_{{\underline{x} [\mu}} B_{{\nu] \underline{x}}}}{G_{{\underline{x}  \underline{x}}}},\label{6.4}\\
V_{{\mu}}&=& \frac{G_{{\mu \underline{x}}} }{G_{{\underline{x} \underline{x}}}},
~~~~~~~~~~~~~~~~~~W_{{\mu}} = B_{{\mu \underline{x}}},\label{6.5}\\
\sigma&=& \frac{1}{2} \log G_{{\underline{x} \underline{x}}}.\label{6.6}
\end{eqnarray}
In order to apply the T-duality rules of KM to our purpose, we will therefore first need to implement
the field redefinitions to go from Hull and Townsend (HT) scheme\footnote{Notice that here we use
the two-loop beta-function equations of the HT scheme which were used in \cite{Wulff}.} \cite{c.hull} to that of KM.
This work has been done in \cite{Wulff}. The field redefinitions that we will use are
\begin{eqnarray}
G_{_{MN}}^{^{(HT)}}&=&G_{_{MN}}^{^{(KM)}} + \alpha'({\cal R}_{_{MN}}-\frac{1}{2}H^2_{_{MN}}),\label{6.7}\\
B_{_{MN}}^{^{(HT)}}&=&B_{_{MN}}^{^{(KM)}} + \alpha'(-H_{_{MNP}}  \nabla^{^{P}}\Phi),\label{6.8}\\
\Phi^{^{(HT)}} &=& \Phi^{^{(KM)}}  + \alpha'(-\frac{3}{32}H^2 + \frac{1}{8}{\cal R}-\frac{1}{2}(\nabla \Phi)^2).\label{6.9}
\end{eqnarray}
Thus, the two-loop T-duality transformation equations in the KM scheme are given by \cite{KM}
\begin{eqnarray}
{\tilde \sigma}&=&-\sigma + \alpha'[(\nabla \sigma)^2 +\frac{1}{8} (e^{2\sigma} Z+e^{-2\sigma} T)],\label{6.10}\\
{\tilde V}_{\mu}&=&W_{\mu} + \alpha'[W_{{\mu \nu}} \nabla^{\nu}\sigma+  \frac{1}{4} h_{\mu\nu\rho} V^{\nu\rho} e^{2\sigma}],\label{6.11}\\
{\tilde W}_{\mu}&=&V_{\mu} + \alpha'[V_{{\mu \nu}} \nabla^{\nu}\sigma-  \frac{1}{4} h_{\mu\nu\rho} W^{\nu\rho} e^{-2\sigma}],\label{6.12}\\
{\tilde b}_{{\mu \nu}} &=& { b}_{{\mu \nu}}  + \alpha'\Big[V_{\rho[\mu} W^{\rho}_{~\nu]}+ (W_{[\mu\rho} \nabla^{\rho}\sigma +\frac{1}{4} e^{2\sigma} h_{[\mu\rho\lambda} V^{\rho \lambda}) V_{\nu]}~~~~~~\nonumber\\
~~~~&&+(V_{[\mu\rho} \nabla^{\rho}\sigma -\frac{1}{4} e^{-2\sigma} h_{[\mu\rho\lambda} W^{\rho \lambda}) W_{\nu]}\Big].\label{6.13}
\end{eqnarray}
These transformations are written using  the following definitions
\begin{eqnarray}
W_{{\mu \nu}}&=&\partial_{\mu} W_\nu -\partial_{\nu} W_\mu,\label{6.14}\\
V_{{\mu \nu}}&=&\partial_{\mu} V_\nu -\partial_{\nu} V_\mu,\label{6.15}\\
h_{\mu\nu\rho}&=& H_{\mu\nu\rho}-3 W_{[\mu\nu} V_{\rho]}.\label{6.16}
\end{eqnarray}
In addition,
\begin{eqnarray}
Z_{{\mu \nu}}&=&V_{{\mu \rho}} V_{\nu}^{~\rho},~~~~~~~~~~~~~~~~~~Z = Z_{\mu}^{~\mu},\label{6.17}\\
T_{{\mu \nu}}&=&W_{{\mu \rho}} W_{\nu}^{~\rho},~~~~~~~~~~~~~~~~T = T_{\mu}^{~\mu}.\label{6.18}
\end{eqnarray}
All the lowering and raising of the indices  will be done with respect to the reduced metric $g_{{\mu \nu}}$.

\subsubsection{Abelian T-duality with respect to the $z$ coordinate up to $\alpha'$-corrections }

In  Sec. \ref{Sec.III} we showed that the $\beta=1$ case of the G\"{o}del spacetimes with
field strength $H_{_{t r \varphi}}=l^2/2$ and a constant dilaton field
satisfy the beta-function equations up to the first order in $\alpha'$
so that the relation between constants $\Lambda$, $l$ and $ \alpha'$
may be expressed as $\Lambda ={1}/{ 2 l^2}+ \alpha'/(4l^4)$.
Here we use the conditions for two-loop conformal invariance of the bosonic
string $\sigma$-model in the HT scheme which were used in \cite{Wulff}.
In the HT scheme, the aforementioned solution can be written as\footnote{To
go from our conventions to those of the HT  \cite{Wulff}  one can send $  \Phi /2 \rightarrow {\Phi}^{^{(HT)}}$
and $2 {H} \rightarrow { H}^{^{(HT)}}$.}
\begin{eqnarray}
{ds^2}^{^{(HT)}} &=&-l^2 d{t}^2 + \frac{l^2 dr^2}{r^2}-2r l^2 ~ dt d\varphi +l^2 d{z}^2,\nonumber\\
H^{^{(HT)}}&=& {l^2} dt \wedge dr \wedge  d\varphi,\nonumber\\
\Phi^{^{(HT)}}&=&\frac{b}{2}.\label{6.27}
\end{eqnarray}
One can use the field redefinitions
in equations \eqref{6.7}-\eqref{6.9} to write the above solution in the KM scheme, expressing
\begin{eqnarray}
{ds^2}^{^{(KM)}} &=&-l^2 d{t}^2 + \frac{l^2 dr^2}{r^2}-2r l^2 ~ dt d\varphi +l^2 d{z}^2 +
\alpha'\Big(\frac{1}{2} d{t}^2 - \frac{dr^2}{2 r^2}+r ~ dt d\varphi\Big),\nonumber\\
H^{^{(KM)}}&=&  l^2~  dt \wedge dr \wedge d\varphi,\nonumber\\
\Phi^{^{(KM)}}  &=&\frac{b}{2} -\frac{3}{8 l^2} \alpha'.\label{6.28}
\end{eqnarray}
Notice that the isometry we want to dualize is that the shift of the $z$ coordinate, i.e., $\underline{x} =z$.
Comparing \eqref{6.28} with equations \eqref{6.1}-\eqref{6.3} or using \eqref{6.4}-\eqref{6.6} one concludes that all components of
$V_{_\mu}$ and $W_{_\mu}$ are zero; consequently, $V_{_{\mu \nu }}=0, W_{_{\mu \nu }}=0, Z=0$ and $T=0$. Furthermore,
\begin{eqnarray}
{\sigma}^{^{(KM)}} = \frac{1}{2} \ln l^2,~~~~~~~~~~
\hat{\phi}   = \Phi^{^{(KM)}}  -\frac{1}{4} \ln l^2.\label{6.29}
\end{eqnarray}
Applying these results to the two-loop T-duality transformation equations \eqref{6.10}-\eqref{6.13}
one can get the dual solution in the KM scheme. Finally, by employing field redefinitions in equations \eqref{6.7}-\eqref{6.9}
we can write the dual solution in the HT scheme, giving
\begin{eqnarray}
{{\tilde ds}^2}^{^{(HT)}} &=&-l^2 d{t}^2 + \frac{l^2 dr^2}{r^2}-2r l^2~ dt d\varphi +\frac{1}{l^2} d{\underline{\tilde x}}^2,\nonumber\\
{\tilde H}^{^{(HT)}}&=&l^2 ~dt \wedge dr \wedge d\varphi,\nonumber\\
{\tilde \Phi}^{^{(HT)}}&=&\frac{b}{2} - \frac{1}{2} \log l^2.\label{6.30}
\end{eqnarray}
It should be noted that utilizing the transformation $\underline{\tilde x}\rightarrow l^2 z$ and shifting
$\frac{b}{2} \rightarrow \frac{b}{2}+ \frac{1}{2} \log l^2$, the dual solution
\eqref{6.30} turns into the original one. Hence, we have shown that
the Abelian T-duals generated by G\"{o}del spacetimes at two-loop order are self-dual.

\subsubsection{Abelian T-duality with respect to the $t$ coordinate up to $\alpha'$-corrections }

Another solution for the two-loop beta-function equations in the HT scheme
including the $\beta=1$ case of the G\"{o}del spacetimes with a non-constant dilaton field is given by
\begin{eqnarray}
{ds^2}^{^{(HT)}} &=&-l^2 d{t}^2 + \frac{l^2 dr^2}{r^2}-2r l^2 ~ dt d\varphi +l^2 d{z}^2,\nonumber\\
H^{^{(HT)}}&=& - l^2~  dt \wedge dr \wedge d\varphi,\nonumber\\
\Phi^{^{(HT)}}&=& a z -\alpha' \frac{z}{16 a l^2},\label{6.19}
\end{eqnarray}
where $4 a^2 =-1+ 2 l^2 \Lambda$. In what follows we want to perform the dualizing on the time coordinate.
Since $G_{{\underline{x} \underline{x}}} =G_{{t t}} $ is negative, we are faced with a timelike case.
In principle, the rules of T-duality are derived assuming that the coordinate to be dualized is spacelike.
Formally, we can achieve that in the timelike case if we first do the analytic continuation $t \rightarrow i t'$,
apply T-duality as usual and then go back doing $t'\rightarrow -i t$.
If one just looks at the NS-NS sector (metric, B-field, dilaton),
this is equivalent to applying the rules without the analytic
continuation, i.e. with $G_{{\underline{x} \underline{x}}}$ negative. The result that he obtains is the same.
In the type $II$ string theory case, the story is more subtle.
As argued in \cite{Hull2}, the timelike T-duals of the $IIA$ and $IIB$ string theories
are new theories that referred to as the $IIB^\ast$ and $IIA^\ast$ string theories,
respectively.

However, by using the Wick rotations $t \rightarrow i t'$ and $\varphi\rightarrow i \varphi'$ the solution \eqref{6.19} is written as
\begin{eqnarray}
{ds^2}^{^{(HT)}} &=&l^2 d{t'}^2 + \frac{l^2 dr^2}{r^2}+2r l^2 ~ dt' d\varphi' +l^2 d{z}^2,\nonumber\\
H^{^{(HT)}}&=&  l^2~  dt' \wedge dr \wedge d\varphi',\nonumber\\
\Phi^{^{(HT)}}&=& a z -\alpha' \frac{z}{16 a l^2}.\label{6.20}
\end{eqnarray}
Accordingly, the isometry we want to dualize is spacelike. That is the shift of the $t'$ coordinate, i.e., $\underline{x} =t'$.
By using the field redefinitions
in equations \eqref{6.7}-\eqref{6.9} one can write the solution \eqref{6.20} in the KM scheme, giving
\begin{eqnarray}
{ds^2}^{^{(KM)}} &=&l^2 d{t'}^2 + \frac{l^2 dr^2}{r^2}+2r l^2 ~ dt' d\varphi' +l^2 d{z}^2 -
\alpha'\Big(\frac{1}{2} d{t'}^2 + \frac{dr^2}{2 r^2}+r ~ dt' d\varphi'\Big),\nonumber\\
H^{^{(KM)}}&=&  l^2~  dt' \wedge dr \wedge d\varphi',\nonumber\\
\Phi^{^{(KM)}}  &=&a z + \Big(\frac{8a^3 -6a-z}{16 a l^2}\Big) \alpha'.\label{6.21}
\end{eqnarray}
Comparing \eqref{6.21} with equations \eqref{6.1}-\eqref{6.3} or using \eqref{6.4}-\eqref{6.6} one obtains that only the nonzero components of
$V_{_\mu}$ and $W_{_\mu}$ are
\begin{eqnarray}
V_{_{\varphi'}} &=&  r,~~~~~~~~~~~W_{_{\varphi'}} = rl^2.\label{6.22}
\end{eqnarray}
In addition,
\begin{eqnarray}
{\sigma}^{^{(KM)}} &=& \frac{1}{2} \log (l^2-\frac{\alpha'}{2})= \frac{1}{2} \log l^2 - \frac{1}{4l^2} \alpha'+ {\cal O}(\alpha'^2),\nonumber\\
\hat{\phi}&=&\Phi^{^{(KM)}} -\frac{1}{4} \log l^2 + \frac{1}{8 l^2} \alpha' +{\cal O}(\alpha'^2).\label{6.23}
\end{eqnarray}
Using the above results, $Z$, $T$, $h_{_{\mu \nu \rho}}$, and only the nonzero components of $V_{_{\mu \nu }}$ and $W_{_{\mu \nu}}$ are derived to be
\begin{eqnarray}
Z &=&  -\frac{2}{l^4} (1-\frac{\alpha'}{2l^2})^{-2},~~~~~~~~~~~T = -2 (1-\frac{\alpha'}{2l^2})^{-2}, \nonumber\\
h_{_{\mu \nu \rho}} &=& 0,~~~~~~V_{_{r \varphi'}} = 1,~~~~~~~~W_{_{r \varphi'}} = l^2.\label{6.24}
\end{eqnarray}
In order to calculate the dual background up to $\alpha'$-corrections one can substitute
equations \eqref{6.22}-\eqref{6.24} into the two-loop T-duality transformation equations \eqref{6.10}-\eqref{6.13}.
Thus, using equations \eqref{6.1}-\eqref{6.3}, the dual solution is obtained in the KM scheme to be of the form
\begin{eqnarray}
{{\tilde ds}^2}^{^{(KM)}} &=&\frac{1}{l^2} d{\underline{{\tilde x}}}^2 + \frac{l^2 dr^2}{r^2}+2r  ~ d\underline{{\tilde x}} d\varphi' +l^2 d{z}^2 -
\alpha'\Big(\frac{1}{2l^4} d{\underline{{\tilde x}}}^2 + \frac{dr^2}{2 r^2}+\frac{r}{l^2} ~ d\underline{{\tilde x}} d\varphi'\Big),\nonumber\\
{\tilde H}^{^{(KM)}}&=&   d\underline{{\tilde x}} \wedge dr \wedge d\varphi',\nonumber\\
{\tilde \Phi}^{^{(KM)}}&=&-\frac{1}{2} \log l^2+ a z + \Big(\frac{8a^3 -6a-z}{16 a l^2}\Big) \alpha'.\label{6.25}
\end{eqnarray}
Finally, one can apply equations \eqref{6.7}-\eqref{6.9} to write the dual solution \eqref{6.25} in the HT scheme. It is then read
\begin{eqnarray}
{{\tilde ds}^2}^{^{(HT)}} &=&\frac{1}{l^2} d{\underline{{\tilde x}}}^2 + \frac{l^2 dr^2}{r^2}+2r  ~ d\underline{{\tilde x}} d\varphi' +l^2 d{z}^2,\nonumber\\
{\tilde H}^{^{(HT)}}&=&d\underline{{\tilde x}} \wedge dr \wedge d\varphi',\nonumber\\
{\tilde \Phi}^{^{(HT)}}&=&-\frac{1}{2} \log l^2+ a z - \frac{z}{16 a l^2} \alpha'.\label{6.26}
\end{eqnarray}
As shown we used the Wick rotations $t \rightarrow i t'$ and $\varphi \rightarrow i \varphi'$, then,
applied the rules of T-duality as usual in order to obtain the dual solution at two-loop order.
Now, one can go back doing the analytic continuation $\underline{{\tilde x}}\rightarrow -i t$ and $\varphi' \rightarrow- i \varphi$.
We represent the results of Abelian T-duality of the G\"{o}del string cosmologies up to $\alpha'$-corrections
in Table 2.

				
                      \begin{table}
						\caption{ Abelian T-duality of the G\"{o}del string cosmologies up to $\alpha'$-corrections}
						\centering
						\begin{tabular}{| p{4em} | p{16em} | p{6.75em} | p{8.75em}|}
							\hline
							\multicolumn{2}{|c|}{\scriptsize  Dualizing with respect to the $z$ coordinate}  & \scriptsize  Conditions for the one-loop solution & \scriptsize Conditions for the two-loop solution\\
							\hline

							\multirow{3}{4em}{\scriptsize Original background}&{\scriptsize $ds^2=l^2\big(-d{t}^2 + \frac{dr^2}{r^2}-2r~ dt d\varphi +d{z}^2\big)$}&  \multirow{3}{6.75em}{\scriptsize $\Lambda=\frac{1}{2l^2}$}   & \multirow{3}{7.25em}{\scriptsize $\Lambda=\frac{1}{2l^2}+\frac{\alpha'}{4l^4}$} \\
							
							&{\scriptsize $H=l^2 dt \wedge dr \wedge d\varphi $} && \\
							
							&{\scriptsize $ \Phi=\frac{b}{2}$}&& \\
							
							\hline

							\multirow{3}{4em}{\scriptsize Dual background}&\scriptsize ${d\tilde {s}}^2=-l^2d{t}^2 + \frac{l^2dr^2}{r^2}-2rl^2~ dt d\varphi +\frac{1}{l^2}d\tilde {\underline{x}}^2$&\multirow{3}{6.75em}{\scriptsize $\tilde\Lambda=\frac{1}{2l^2}$}   &\multirow{3}{7.25em}{\scriptsize $\tilde \Lambda=\frac{1}{2l^2}+\frac{\alpha'}{4l^4}$} \\

							&{\scriptsize $\tilde H=l^2 dt \wedge dr \wedge d\varphi $}& & \\
							&\scriptsize $\tilde\Phi=\frac{b}{2}-\frac{1}{2}\log{l^2}$ & & \\
							\hline
						\end{tabular}
						\begin{tabular}{| p{4em} | p{16em} | p{6.75em} | p{8.75em}|}
							\hline
							\multicolumn{2}{|c|}{\scriptsize  Dualizing with respect to the $t$ coordinate}  & \scriptsize  Conditions for the one-loop solution & \scriptsize Conditions for the two-loop solution\\
							\hline

							\multirow{2}{4em}{\scriptsize Original background}&{\scriptsize $ds^2=l^2\big(-d{t}^2 + \frac{dr^2}{r^2}-2r~ dt d\varphi +d{z}^2\big)$}&  \scriptsize $ \Phi=az,$   & \scriptsize $ \Phi=az-\alpha'\frac{z}{16al^2},$ \\
							
							&{\scriptsize $H=-l^2 dt \wedge dr \wedge d\varphi $} &\scriptsize $4a^2=2l^2\Lambda-1$& \scriptsize $4a^2=2l^2\Lambda-1$\\

							\hline

							\multirow{2}{4em}{\scriptsize Dual background}&\scriptsize ${d\tilde {s}}^2=-\frac{1}{l^2}d\tilde{\underline x}^2 + \frac{l^2dr^2}{r^2}-2r~ d\tilde{\underline x} d\varphi +l^2dz^2$&\scriptsize $ \tilde\Phi=-\frac{1}{2}\log{l^2} +az,$    &\scriptsize $ \tilde\Phi=-\frac{1}{2}\log{l^2}+az-\alpha'\frac{z}{16al^2},$ \\

							&{\scriptsize $\tilde H= dr \wedge d\tilde{\underline x} \wedge d\varphi $}&  \scriptsize $4a^2=2l^2\tilde\Lambda-1$& \scriptsize $4a^2=2l^2\tilde\Lambda-1$ \\
							
							\hline
						\end{tabular}
					\end{table}


\section{\label{Sec.VII} Non-Abelian T-dualization of G\"{o}del string cosmologies}
We now wish to apply the formulation of Sec. \ref{Sec.IV} in order to study the non-Abelian T-dualization of the G\"{o}del spacetimes.
In this section, we explicitly construct some of the PL T-dual $\sigma$-models on $2+2$- and $3+1$-dimensional target
manifolds. In any case, the metrics of the original $\sigma$-models represent the G\"{o}del spacetimes.
From the analysis of Sec. \ref{Sec.III}, it is followed that the backgrounds of the models are conformally invariant up to two-loop order such that
they satisfy Eqs. \eqref{a.2}-\eqref{a.4}. We clarify the corresponding dual spacetimes structure, as well as
the conformal invariance conditions of them.
In a particular case of the models, we study the PL self-duality of $AdS_3 \times \mathbb{R}$ space.

\subsection{\label{Sec.VI.i} Non-Abelian T-duality from a $2+2$-dimensional manifold with the $A_2$ Lie group}

As discussed in Ref. \cite{Eghbali.2}, the $2+2$-dimensional manifold ${\cal M} \approx O \times  G$ with two-dimensional
real non-Abelian Lie group ${ G}=A_2$ is wealthy.
We shall obtain the G\"{o}del spacetimes from a  T-dualizable $\sigma$-model
constructed out of the $2+2$-dimensional manifold ${\cal M} \approx O \times  G$ where $ G$  acts freely on
${\cal M}$, while $O$ is the orbit of ${ G}$ in $\mathcal{M}$.
In this way, the non-Abelian T-duality of the G\"{o}del spacetimes is studied here.
The dual Lie group ${\tilde { G}}$ acting freely
on the dual manifold ${\tilde {\cal M}} \approx O \times {\tilde { G}}$ is considered to be two-dimensional Abelian Lie group $2 A_1$.
The Lie algebras of the Lie groups ${ G}$ and $\tilde { G}$ are denoted by
${\cal A}_2$ and $2{\cal A}_1$, respectively.
As mentioned in Sec. \ref{Sec.IV},  having  Drinfeld doubles one can construct PL T-dual $\sigma$-models on them.
The four-dimensional Lie algebra of the Drinfeld double $({\cal A}_2 , 2{\cal A}_1)$ is defined by the following nonzero commutation relations:
\begin{eqnarray}\label{VI.1}
[T_1 , T_2]~=~T_2,~~[T_1 ~, ~{\tilde T}^2]=-{\tilde T}^2,~~[T_2 ~, ~{\tilde T}^2]={\tilde T}^1,
\end{eqnarray}
where $(T_1 , T_2)$ and $({\tilde T}^1 , {\tilde T}^2)$ are the basis of ${\cal A}_2$ and $2{\cal A}_1$, respectively.

In order to write the action of the $\sigma$-models \eqref{c.5} and \eqref{c.11}
on the $2+2$-dimensional manifolds ${\cal M}$ and $\tilde {\cal M}$ we need to calculate the components of the right-invariant Maurer-Cartan
forms $R_{\pm}^a$ and ${\tilde R}_{\pm_{a}}$ on the Lie groups $A_2$ and $2A_1$, respectively.
To this end, we parametrize elements of the $A_2$ and $2A_1$ as
\begin{eqnarray}\label{VI.2}
g~=~e^{x_{_1} T_1}~e^{x_{_2} T_2},~~~~~~~~\tilde g~=~e^{{\tilde x}_{_1} {\tilde T}^1}~e^{{\tilde x}_{_2} {\tilde T}^2},
\end{eqnarray}
where $(x_{_1}, x_{_2})$ and $({\tilde x}_{_1}, {\tilde x}_{_2})$ stand for the coordinates
of the Lie groups $A_2$ and $2A_1$, respectively. $R_{\pm}^a$'s and ${\tilde R}_{\pm_{a}}$'s  are then obtained to be of the forms
\begin{eqnarray}
R_{\pm}^1&=& \partial_{\pm} x_{_1},~~~~~~~~R_{\pm}^2~=~ e^{x_{_1}}~\partial_{\pm} x_{_2},\label{VI.3}\\
{\tilde R}_{\pm_{1}}&=& \partial_{\pm} {\tilde x}_{_1},~~~~~~~~{\tilde R}_{\pm_{2}} = \partial_{\pm} {\tilde x}_{_2}.\label{VI.4}
\end{eqnarray}
For our purpose it is also necessary to compute the matrices $\Pi^{^{ab}}(g)$ and ${\tilde \Pi}_{_{ab}}(\tilde g)$.
Using Eqs. \eqref{VI.1} and \eqref{VI.2} and also applying Eq. \eqref{c.4} for both the Lie groups $A_2$ and $2A_1$, we get
\begin{eqnarray}\label{VI.5}
\Pi^{^{ab}}(g) ~=~0,~~~~~~~~~~~~{\tilde \Pi}_{_{ab}}(\tilde g)\;=\;\left( \begin{array}{cc}
                     0 & -{\tilde x}_{_2}\\

                     {\tilde x}_{_2} & 0
                      \end{array} \right).
\end{eqnarray}
Let us now choose the spectator-dependent background  matrices as
{\begin{eqnarray}
E_{0\;ab}&=&\left( \begin{array}{cc}
                    l^2 ~& 0\\
                    0  ~ & l^2(\beta -1)
                      \end{array} \right),~~~~~~~~~
F^{^{(1)}}_{a j}=\left( \begin{array}{cc}
                    0 ~& 0\\
                    2\mathbb{E}-l^2 ~& 2\mathbb{N}
                      \end{array} \right),\nonumber\\
F^{^{(2)}}_{i b}&=&\left( \begin{array}{cc}
                    0~ & -(2\mathbb{E}+l^2)\\
                    0 ~& -2\mathbb{N}
                      \end{array} \right),~~~~~~~~~F_{ij}=\left( \begin{array}{cc}
                    -l^2 ~& 0\\
                    0 ~& l^2
                      \end{array} \right).\label{VI.6}
\end{eqnarray}}
Inserting \eqref{VI.6} into Eqs. \eqref{c.6}-\eqref{c.9} and then
utilizing formulas \eqref{VI.3} and \eqref{VI.5} together with \eqref{c.5},
the original $\sigma$-model is worked out to be
\begin{eqnarray}
S &=& \frac{1}{2} \int d \sigma^+ d \sigma^-~\Big[l^2 \partial_+ x_{_1} \partial_- x_{_1}+ l^2(\beta -1) e^{2x_{_1}} \partial_+ x_{_2} \partial_- x_{_2}
+ (2\mathbb{E}-l^2) e^{x_{_1}} \partial_+ x_{_2} \partial_- y_{_1} \nonumber\\
&&~~~~~~~~~~~~~~~~~~-(2\mathbb{E}+l^2) e^{x_{_1}} \partial_+ y_{_1}   \partial_- x_{_2}
+2\mathbb{N}  e^{x_{_1}} (\partial_+ x_{_2}  \partial_- y_{_2} - \partial_+ y_{_2} \partial_- x_{_2})
 \nonumber\\
&&~~~~~~~~~~~~~~~~~~-l^2 \partial_+ y_{_1} \partial_- y_{_1}+ l^2 \partial_+ y_{_2} \partial_- y_{_2}\Big]-\frac{1}{4 \pi} \int d\sigma^{+}  d\sigma^{-} ~R^{^{(2)}} \Phi(X),\label{VI.7}
\end{eqnarray}
where $y^i =(y_{_1} , y_{_2})$ are the  coordinates of the orbit $O$ of $G$ in manifold ${\cal M}$.
To be more specific, we use the coordinate transformation
\begin{eqnarray}\label{VI.8}
e^{x_{_1}} = r,~~~~{x_{_2}} = \varphi,~~~~~~y_{_1} = t, ~~~~~~~y_{_2} = z.
\end{eqnarray}
Then, by identifying the action \eqref{VI.7}  with the $\sigma$-model of the form \eqref{a.1},
the corresponding line element and antisymmetric tensor  can be, respectively, cast in the forms
\begin{eqnarray}
ds^2&=&l^2\Big[-d{t}^2 + \frac{dr^2}{r^2}+(\beta-1)r^2~ d{\varphi}^2-2r~ dt d\varphi +d{z}^2\Big],\label{VI.9}\\
{ B} &=& 2\mathbb{E} r d\varphi \wedge dt + 2\mathbb{N} r d\varphi \wedge dz.\label{VI.10}
\end{eqnarray}
The line element \eqref{VI.9} is nothing but the G\"{o}del metric \eqref{b.10}, and
the field strength corresponding to the $B$-field \eqref{VI.10} is easily obtained to be the same form as \eqref{b.17}.
Because of the invertibility of the matrix $E_{0}$,
the parameter $\beta$ must be here considered to be different from one. Accordingly, the model \eqref{VI.7}
will not be conformally invariant  at one-loop level.
As discussed in Sec. \ref{Sec.III} (class B of solutions), the metric \eqref{VI.9} and field strength \eqref{b.17}
together with a constant dilaton field
satisfy the two-loop beta-function equations \eqref{a.2}-\eqref{a.4}.

In order to construct the dual $\sigma$-model corresponding to the model \eqref{VI.7} we use the action \eqref{c.11}.
The dual coupling matrices can be obtained by
inserting \eqref{VI.5} and \eqref{VI.6} into \eqref{c.12}-\eqref{c.15}.
They are then read
{\begin{eqnarray}\nonumber
{\tilde E}^{ab}=\frac{1}{\Delta}\left( \begin{array}{cc}
                    l^2(\beta-1) & {\tilde x}_{_2}\\
                   -{\tilde x}_{_2} & l^2
                      \end{array} \right),~~~~~~~
{\tilde \phi}_{_{ij}} =\left( \begin{array}{cc}
                    -l^2\big[1-\frac{1}{\Delta}(4\mathbb{E}^2-l^4)\big] & \frac{2\mathbb{N}l^2}{\Delta}(2\mathbb{E}+l^2)\\
                    \frac{2\mathbb{N}l^2}{\Delta}(2\mathbb{E}-l^2) & l^2\big(1+\frac{4\mathbb{N}^2}{\Delta}\big)
                      \end{array} \right),
\end{eqnarray}}
\vspace{-2mm}
{\begin{eqnarray}\label{VI.12}
{\tilde \phi}^{^{(1)^{ a}}}_{~~~j} =\frac{1}{\Delta}\left( \begin{array}{cc}
                   (2\mathbb{E}-l^2){{\tilde x_{_2}}}~ & 2\mathbb{N} {{\tilde x_{_2}}}\\
                    (2\mathbb{E}-l^2) l^2~ & 2\mathbb{N}l^2
                      \end{array} \right),~~
{{\tilde \phi}^{^{{(2)}^{b}}}}_{~i} =\frac{1}{\Delta}\left( \begin{array}{cc}
                   -(2\mathbb{E}+l^2){{\tilde x_{_2}}}~ & (2\mathbb{E}+l^2) l^2 \\
                    -2\mathbb{N} {{\tilde x_{_2}}}~     & 2\mathbb{N}l^2
                      \end{array} \right),
\end{eqnarray}}
where $\Delta={\tilde x_{_2}}^2+(\beta-1)l^4 $.
Finally, one can write down the action of dual $\sigma$-model in the following form
\begin{eqnarray}\label{VI.13}
\tilde S &=& \frac{1}{2} \int d\sigma^{+}  d\sigma^{-}\Big\{-l^2(1-\frac{4\mathbb{E}^2-l^4}{\Delta})\partial_{+} t  \partial_{-} t+
l^2(1+\frac{4\mathbb{N}^2}{\Delta})\partial_{+} z  \partial_{-} z \nonumber\\
&&+\frac{1}{\Delta} \Big[l^2(\beta-1)\partial_{+} {\tilde x_{_1}} \partial_{-} {\tilde x_{_1}} +l^2 \partial_{+} {\tilde x_{_2}} \partial_{-} {\tilde x_{_2}} + {\tilde x_{_2}}  (\partial_{+} {\tilde x_{_1}} \partial_{-} {\tilde x_{_2}}-\partial_{+} {\tilde x_{_2}} \partial_{-} {\tilde x_{_1}})\nonumber\\
&&+2\mathbb{N}{\tilde x_{_2}} (\partial_{+} {\tilde x_{_1}} \partial_{-} z -
\partial_{+} z  \partial_{-} {\tilde x_{_1}})+2\mathbb{N}l^2 (\partial_{+} {\tilde x_{_2}} \partial_{-} z +\partial_{+} z  \partial_{-} {\tilde x_{_2}})
+(2\mathbb{E}-l^2) ({\tilde x_{_2}} \partial_{+} {\tilde x_{_1}} \partial_{-} t\nonumber\\
&&+l^2\partial_{+} {\tilde x_{_2}}  \partial_{-} t+2\mathbb{N}l^2
\partial_{+} z \partial_{-} t)+(2\mathbb{E}+l^2) (-{\tilde x_{_2}} \partial_{+}t  \partial_{-}{\tilde x_{_1}} +l^2\partial_{+} t   \partial_{-}  {\tilde x_{_2}}+2\mathbb{N}l^2
\partial_{+} t\partial_{-} z)\Big]\Big\}\nonumber\\
&&~~~~~~~~~~~~~~~~~~~~~~~~~~~~~~-\frac{1}{4 \pi} \int d\sigma^{+}  d\sigma^{-} ~R^{^{(2)}}  {\tilde \Phi({\tilde X})}.
\end{eqnarray}
Comparing  the above action with the  $\sigma$-model action of the form \eqref{a.1},
the line element and the tensor
field ${\tilde B}$ take the following forms
\begin{eqnarray}
d\tilde s^2&=&\frac{1}{\Delta}\Big[-l^2({\tilde x_{_2}}^2 +\beta l^4 -4\mathbb{E}^2) d{t}^2 +
l^2\big({\tilde x_{_2}}^2 +(\beta-1) l^4 +4\mathbb{N}^2\big) d{z}^2 +l^2(\beta-1) d {\tilde x_{_1}}^2 \nonumber\\
&&~~~~~~~~~~~~+l^2 d {\tilde x_{_2}}^2 -2 l^2  {\tilde x_{_2}}
d {\tilde x_{_1}} dt + 4\mathbb{E}l^2 d {\tilde x_{_2}} dt+ 4\mathbb{N}l^2 d {\tilde x_{_2}} dz +8\mathbb{E}\mathbb{N}l^2 d z dt\Big],\label{VI.14}\\
{ \tilde B} &=& \frac{1}{\Delta}\Big[{\tilde x_{_2}} d {\tilde x_{_1}} \wedge  d {\tilde x_{_2}} + 2\mathbb{E} {\tilde x_{_2}} d {\tilde x_{_1}} \wedge  dt
 + 2\mathbb{N} {\tilde x_{_2}} d {\tilde x_{_1}} \wedge  dz  -l^4  d {\tilde x_{_2}} \wedge  dt
- 2\mathbb{N} l^4  d z \wedge  dt\Big].~~~~~~~\label{VI.15}
\end{eqnarray}
In order to enhance and clarify the structure of the dual spacetime,
we can test whether there are true singularities
by calculating the scalar curvature, which is
\begin{eqnarray}
{\tilde {\cal R}} &=&  \frac{1}{2\beta l^2{\Delta^2}} \Big[(1-4 \beta) {\tilde x_{_2}}^4
+\Big(4\mathbb{N}^2 (1-20 \beta) +2(\beta -1)(l^4+40\mathbb{E}^2)\Big)  {\tilde x_{_2}}^2  \nonumber\\
&&~~~~~~-32\mathbb{E}^2 l^4 (\beta -1)^2 +4 \mathbb{N}^2 l^4 (\beta -1)  (1+8 \beta) +l^8 (\beta -1)^2 (1+4 \beta)\Big].\label{VI.16}
\end{eqnarray}
As it is seen from the formulas \eqref{VI.14} and \eqref{VI.16}, the
regions ${\tilde x_{_2}} =\pm  l^2\sqrt{(1-\beta)}$ are true curvature singularities.
Therefore, the dual metric has true  singularities for the range $0 < \beta < 1$.
The results indicate that the dual spacetime corresponding to the G\"{o}del metrics allowing CTC's has true singularities,
because the metric \eqref{b.10} allows CTC's for the range $0 < \beta<1$ as mentioned in Sec. \ref{Sec.III}.

In order to investigate the conformal invariance conditions of the dual model
up to the first order in  $\alpha'$, we look at the two-loop beta-function equations.
Before proceeding to this, one easily gets that
the field strength corresponding to the ${\tilde B}$-field \eqref{VI.15} is
\begin{eqnarray}
{\tilde H} &=&   \frac{1}{\Delta^2} \Big[\mathbb{E}\big({\tilde x_{_2}}^2-(\beta-1)l^4\big) d {\tilde x_{_1}} \wedge d {\tilde x_{_2}}\wedge dt
+\mathbb{N}\big({\tilde x_{_2}}^2-(\beta-1)l^4\big) d {\tilde x_{_1}} \wedge d {\tilde x_{_2}}\wedge dz\nonumber\\
&&~~~~~~~~~~~~~~~~~~~~~~~~~~~~~~~~~~~~~~~~~~~~~~~~~~~~~~~~-2\mathbb{N}l^4 {\tilde x_{_2}}  d {\tilde x_{_2}} \wedge d t\wedge dz\Big].\label{VI.17}
\end{eqnarray}
Now by solving the field equations \eqref{a.2}-\eqref{a.4}
for the metric \eqref{VI.14} and the field strength \eqref{VI.17}
one concludes that there is no suitable dilaton field
to satisfy these equations. In the next subsection, we study
the non-Abelian T-duality of the G\"{o}del spacetimes by constructing a pair of the mutually dual $\sigma$-models  on
 $3+1$-dimensional manifolds with the $A_2 \oplus A_1$ and $3 A_1$  Lie groups.
It is then shown the T-dual $\sigma$-models that are conformal at one-loop remain conformal at two-loop.

\subsection{\label{Sec.VI.iv}  Non-Abelian T-duality from a $3+1$-dimensional manifold with the $A_2 \oplus A_1$  Lie group}

There is a possibility that we can also get the G\"{o}del spacetimes from a T-dualizable $\sigma$-model constructing on a $3+1$-dimensional manifold
${\cal M} \approx O \times  G$, in which $ G$  is
three-dimensional decomposable Lie group $A_2 \oplus A_1$ \footnote{Notice that the Lie algebra
${\cal A}_2 \oplus {\cal A}_1$ of the Lie group $A_2 \oplus A_1$ is isomorphic to the Lie algebra of Bianchi type III.} acting freely on ${\cal M}$, while
the orbit $O$ is, here, a one-dimensional space with time coordinate  $y^{i}=\{t\}$.
The dual model is constructed on manifold $\tilde {{\cal M}} \approx O \times \tilde {G}$,
where $\tilde {G}$ is three-dimensional Abelian Lie group $3 A_1$.
The Lie algebra of the semi-Abelian double $({\cal A}_2 \oplus {\cal A}_1 , 3{\cal A}_1)$
is generated by the generators $(T_a , {\tilde T}^a)$ with the following  Lie brackets \cite{{JR},{Snobl},{RHR}}
\begin{eqnarray}\label{VI.18}
[T_1 , T_2] &=& T_2,~~~~[T_1 ~, ~{\tilde T}^2]=-{\tilde T}^2,~~~~{[T_2 ~, ~{\tilde T}^2]} = {\tilde T}^1,~~~~[T_3~ , ~.]~=~0,\nonumber\\
{[{\tilde T}^1 ~, ~.] }&=& 0,\hspace{7mm}[{\tilde T}^3 ~, ~.]=0.
\end{eqnarray}
Taking a convenient element of the Lie group $A_2 \oplus A_1$,
\begin{eqnarray}\label{VI.19}
g=e^{x_{_1} T_{_1}}~e^{x_{_2} T_{_2}}~e^{x_{_3} T_{_3}},
\end{eqnarray}
where $(x_{_1}, x_{_2}, x_{_3})$ stand for the coordinates of $A_2 \oplus A_1$,
we immediately find that $R_{\pm}^1= \partial_{\pm} x_{_{1}}$,
$R_{\pm}^2= e^{x_{_{1}}} \partial_{\pm} x_{_{2}}$ and $R_{\pm}^3= \partial_{\pm} x_{_{3}}$.
In addition, we parameterize the Lie group $3A_1$ with
coordinates $({\tilde x}_{_1}, {\tilde x}_{_2} , {\tilde x}_{_3})$ so that its element
is defined as in \eqref{VI.19} by replacing untilded quantities with tilded ones.  Hence,
using \eqref{c.4} with and without tilded quantities we obtain
\begin{eqnarray}\label{VI.20}
\Pi^{^{ab}}(g)\;=\; 0,~~~~~~~~~~{\tilde \Pi}_{_{ab}}(\tilde g)\;=\;\left( \begin{array}{ccc}
                     0 & -{\tilde x}_{_2}&0\\

                     {\tilde x}_{_2} & 0 &0\\

                     0 & 0& 0
                      \end{array} \right).
\end{eqnarray}
The coupling matrices of the original $\sigma$-model are, here, obtained by considering the spectator-dependent background  matrices
{\small \begin{eqnarray}\label{VI.21}
E_{0\;ab} =\left( \begin{array}{ccc}
                    l^2  & s & 0\\
                       -s & (\beta-1)l^2 & 2\mathbb{N}\\
                       0 & -2\mathbb{N} & l^2
                      \end{array} \right),~F^{^{(1)}}_{a j}=\left( \begin{array}{c}
                    0 \\
                    2\mathbb{E}-l^2 \\
                    0
                      \end{array} \right),~F^{^{(2)}}_{i b}=\left( \begin{array}{ccc}
                    0~ & -(2\mathbb{E}+l^2) ~&0
                      \end{array} \right),
\end{eqnarray}}
and $ F_{ij} =-l^2$ for some constant $s$. Inserting \eqref{VI.20} and \eqref{VI.21}
into Eqs. \eqref{c.6}-\eqref{c.9}, and using \eqref{c.5}, then  we obtain the original $\sigma$-model in the following form
\begin{eqnarray}
S &=& \frac{1}{2} \int d \sigma^+ d \sigma^-~\Big[l^2 \partial_+ x_{_1} \partial_- x_{_1}+ l^2(\beta -1) e^{2x_{_1}} \partial_+ x_{_2} \partial_- x_{_2}
+ (2\mathbb{E}-l^2) e^{x_{_1}} \partial_+ x_{_2} \partial_- t \nonumber\\
&&~~~~~~~~~~~~~~~~~~-(2\mathbb{E}+l^2) e^{x_{_1}} \partial_+ t  \partial_- x_{_2}
+2\mathbb{N}  e^{x_{_1}} (\partial_+ x_{_2}  \partial_- x_{_3} - \partial_+ x_{_3} \partial_- x_{_2})
 \nonumber\\
&&~~~~~~~~~~~~~~~~~~-l^2 \partial_+ t \partial_- t+ l^2 \partial_+ x_{_3} \partial_- x_{_3}+s  e^{x_{_1}} (\partial_+ x_{_1}  \partial_- x_{_2} - \partial_+ x_{_2} \partial_- x_{_1})\Big] \nonumber\\
&&~~~~~~~~~~~~~~~~~~-\frac{1}{4 \pi} \int d\sigma^{+}  d\sigma^{-} ~R^{^{(2)}} \Phi(X).\label{VI.22}
\end{eqnarray}
Utilizing the coordinate transformation
$e^{x{_{_1}}}=r, x{_{_2}} =\varphi, x{_{_3}} =z$,
the metric of the model becomes  the same as the G\"{o}del metric \eqref{b.10}, and $B$-field is given by
\begin{eqnarray}
B=2\mathbb{E} r d\varphi \wedge dt + 2\mathbb{N} r d\varphi \wedge dz + s  d r  \wedge d\varphi,\label{VI.23}
\end{eqnarray}
such that the corresponding field strength takes the same form as \eqref{b.17}.
One can check that only the case of $\beta=1$ of the model \eqref{VI.22} with conditions $\mathbb{E}^2=l^4/4$ and $\mathbb{N}=0$ is
conformally invariant up to zeroth order in $\alpha'$. The dilaton filed that makes the model conformal is obtained by
using equation \eqref{c.16.1} to be $ \Phi = \phi^{^{(0)}}+x_{_1}= \phi^{^{(0)}}+\log r$ such that $\phi^{^{(0)}}$
must be $\phi^{^{(0)}}=f z -\log r+b$. Finally we get
\begin{eqnarray}
\Phi = f z +b,\label{VI.23.1}
\end{eqnarray}
where $f^2=-1+ 2 l^2 \Lambda $.

In order to obtain the dual $\sigma$-model
for \eqref{VI.22}, we use the action
\eqref{c.11}. The dual coupling matrices can be obtained by
inserting \eqref{VI.20} and \eqref{VI.21} into \eqref{c.12}-\eqref{c.15}. Finally,
the dual background is read
\begin{eqnarray}
d\tilde s^2&=&\frac{1}{\hat{\Delta}}\Big\{-l^2\big[(s-{\tilde x_{_2}})^2 +\beta l^4 +4(\mathbb{N}^2-\mathbb{E}^2)\big] d{t}^2
 +\frac{1}{l^2} \big[4\mathbb{N}^2 +l^4(\beta-1)\big] d {\tilde x_{_1}}^2 \nonumber\\
&&~~~~~~~~~+l^2 d {\tilde x_{_2}}^2+\frac{1}{l^2} \big[(s-{\tilde x_{_2}})^2+ l^4(\beta-1)\big]
d {\tilde x_{_3}}^2 -4\mathbb{N} l^2  d {\tilde x_{_3}} dt + 4\mathbb{E}l^2 d {\tilde x_{_2}} dt\nonumber\\
&&~~~~~~~~~+2 l^2  (s-{\tilde x_{_2}}) d {\tilde x_{_1}} dt+\frac{4}{l^2} \mathbb{N} (s-{\tilde x_{_2}}) d {\tilde x_{_1}} d {\tilde x_{_3}} \Big\},\label{VI.24}\\
{ \tilde B} &=& \frac{1}{\hat{\Delta}}\Big\{({\tilde x_{_2}}-s) d {\tilde x_{_1}} \wedge  d {\tilde x_{_2}} + 2\mathbb{E} ({\tilde x_{_2}}-s) d {\tilde x_{_1}} \wedge  dt
 - 2\mathbb{N} d {\tilde x_{_2}} \wedge  d {\tilde x_{_3}}
 \nonumber\\
&&~~~~~~~~~~~~~~~~~~~~~~~~~~~~~~~~~~~~~~~-l^4  d {\tilde x_{_2}} \wedge  dt+4\mathbb{E}\mathbb{N}  d {\tilde x_{_3}} \wedge  dt\Big\},~~~~\label{VI.25}
\end{eqnarray}
where $\hat{\Delta}=(s-{\tilde x_{_2}})^2 +4\mathbb{N}^2+(\beta-1)l^4 $.
One immediately gets that the only nonzero components of the field strength corresponding to the
$\tilde B$-field  \eqref{VI.25} are
\begin{eqnarray}
{\tilde H}_{_{{\tilde x_{_1}} {\tilde x_{_2}} t}} = \frac{\mathbb{E}}{{\hat{\Delta}}^2} \big[(s-{\tilde x_{_2}})^2 -4\mathbb{N}^2-(\beta-1)l^4\big],
  ~~~~~~~{\tilde H}_{_{{\tilde x_{_2}} {\tilde x_{_3}} t}} =\frac{4\mathbb{E}\mathbb{N}(s-{\tilde x_{_2}})}{{\hat{\Delta}}^2}. \label{VI.26}
\end{eqnarray}
Before proceeding to investigate the conformal invariance conditions of the dual background,
let us enhance and clarify the structure of the dual spacetime. Similar to the original model we
impose the conditions $\beta=1,~\mathbb{E}^2={l^4}/{4}$ and $\mathbb{N}=0$
on the dual solution (equations \eqref{VI.24} and \eqref{VI.25}),
then by shifting ${\tilde x_{_2}}$ to ${\tilde x_{_2}} +s$ we get
\begin{eqnarray}
d\tilde s^2&=&-l^2 d{t}^2 + \frac{1}{l^2} d {\tilde x_{_3}}^2 +\frac{l^2}{{\tilde x_{_2}}^2} d {\tilde x_{_2}}^2 -\frac{2l^2}{{\tilde x_{_2}}} \big(d {\tilde x_{_1}} dt
-\frac{l^2}{{\tilde x_{_2}}} d {\tilde x_{_2}} dt\big),\label{VI.29}\\
{ \tilde B} &=& \frac{1}{{\tilde x_{_2}}} \Big[d {\tilde x_{_1}} \wedge  d {\tilde x_{_2}} + l^2  d {\tilde x_{_1}} \wedge  dt
-\frac{l^4}{{\tilde x_{_2}}} d {\tilde x_{_2}} \wedge  dt\Big].\label{VI.30}
\end{eqnarray}
Here we have used $\mathbb{E}=l^2/2$.
To have a better understanding of the structure of metric \eqref{VI.29},
we use the coordinate transformation
\begin{eqnarray}
{\tilde x_{_1}} ~=~-l^2 \ln r + \varphi, ~~~~~~~{\tilde x_{_2}} ~=~\frac{1}{r},~~~~~~~{\tilde x_{_3}} = l^2 z.\label{VI.31}
\end{eqnarray}
Accordingly, the metric \eqref{VI.29} turns into the case of $\beta=1$ of the G\"{o}del metric \eqref{b.10}\footnote{Notice that
the G\"{o}del metric \eqref{b.10} with the condition $\beta=1$ represents the $AdS_3 \times \mathbb{R}$ space locally as shown in Eq. \eqref{b.11.2}.}.
This particular case of the dual solution is conformally invariant
up to the one-loop order with a new dilaton field which is obtained by using equations
\eqref{c.16.2} and \eqref{VI.23.1}, giving
\begin{eqnarray}
{\tilde \Phi} ~=~{\tilde f}{\tilde x_{_3}}+{\tilde b} = {\tilde f} l^2 z+{\tilde b}.\label{VI.31.1}
\end{eqnarray}
where ${\tilde f}^2=\big(2l^2 \Lambda -1\big)/l^4 $.
It should be noted that the dual background given by equations \eqref{VI.29} and \eqref{VI.30} is also conformally invariant
up to two-loop order with the same dilaton filed \eqref{VI.31.1}.
After performing the transformation \eqref{VI.31} on the ${\tilde B}$-field \eqref{VI.30}, the only nonzero component of
the corresponding field strength becomes ${\tilde H}_{_{t r  \varphi}} = l^2 /2$. This
result is in agreement with the case $(1')$ of the solutions of class A.
{\it Thus, we showed that the $AdS_3 \times \mathbb{R}$ space does remain invariant under the non-Abelian T-duality transformation, that is, the model is PL self-dual.} If one uses the conditions $\beta=1,~\mathbb{N}=0$ and $\mathbb{E}=-l^2/2$, then, they will obtain similar results.

\subsection{\label{Sec.VI.v} Non-Abelian T-duality from a $3+1$-dimensional manifold with  the $SL(2 , \mathbb{R})$ Lie group}

In the preceding subsection, we studied the non-Abelian T-duality of the G\"{o}del spacetimes by applying  the $3+1$-dimensional manifold
${\cal M} \approx O \times  G$ with $G= A_2 \oplus A_1$. In addition, we can only derive the case of $\beta =1$ of the
G\"{o}del metrics from a $3+1$-dimensional manifold with  the $SL(2 , \mathbb{R})$ Lie group and then obtain
the corresponding dual spacetime. In this way, the pair of the mutually dual models are constructed on semi-Abelian double
$(sl(2, \mathbb{R}) , 3{\cal A}_1)$.
The Lie algebra of  double $(sl(2, \mathbb{R}) , 3{\cal A}_1)$
is defined by nonzero Lie brackets as \cite{{JR},{Snobl},{RHR}},
\begin{eqnarray}\label{VI.32}
[T_1 , T_2] &=& T_2,~~~~~~[T_1 , T_3] = -T_3,~~~~~~~~~[T_2 , T_3] = 2T_1,~~~~~~
[T_1 ~, ~{\tilde T}^2]=-{\tilde T}^2,\nonumber\\
{[T_1 ~, ~{\tilde T}^3] }&=& {\tilde T}^3,~~~~~[T_2 ~, ~{\tilde T}^1] =-2 {\tilde T}^3,~~~~~[T_2 ~, ~{\tilde T}^2] = {\tilde T}^1,~~~~~[T_3 ~, ~{\tilde T}^3] =- {\tilde T}^1,\nonumber\\
{[T_3 ~, ~{\tilde T}^1] }&=& -2{\tilde T}^2.
\end{eqnarray}
where $(T_1 , T_2, T_3)$ and $({\tilde T}^1 , {\tilde T}^2, {\tilde T}^3)$ are the basis of the Lie algebras $sl(2, \mathbb{R})$ and $3{\cal A}_1$, respectively.
We note that  the double $(sl(2, \mathbb{R}) , 3{\cal A}_1)$ has the vanishing trace in the adjoint representations.
In such a situation, at the one-loop level a conformally invariant $\sigma$-model leads, under the PL
duality, to a dual theory with the same property \cite{N.Mohammedi}.
Now we parametrize an element of $SL(2, \mathbb{R})$ as
\begin{eqnarray}\label{VI.33}
g~=~e^{x_{_{2}} T_2}~e^{x_{_{1}} T_1} ~e^{x_{_{3}} T_3},
\end{eqnarray}
where $(x_{_{1}}, x_{_{2}}, x_{_{3}})$ are the coordinates of $SL(2, \mathbb{R})$. Then one gets
the corresponding one-forms components in the following way
\begin{eqnarray}\label{VI.34}
R_{\pm}^1&=& \partial_{\pm} x_{_{1}} + 2e^{-x_{_{1}}} x_{_{2}} \partial_{\pm} x_{_{3}},\nonumber\\
R_{\pm}^2&=& -x_{_{2}} \partial_{\pm} x_{_{1}} + \partial_{\pm} x_{_{2}} - {x_{_{2}}}^2 e^{-x_{_{1}}}  \partial_{\pm} x_{_{3}},\nonumber\\
R_{\pm}^3&=& e^{-x_{_{1}}} \partial_{\pm} x_{_{3}}.
\end{eqnarray}
Since the dual Lie group is considered to be Abelian,
hence, by using \eqref{c.4} and \eqref{VI.32}  it is followed that
$\Pi(g) =0$. Using the above results and also choosing the spectator-dependent matrices in the form
\begin{eqnarray}\label{VI.35}
E_{0\;ab} =\left( \begin{array}{ccc}
                    \frac{l^2}{4}  & 0 & 0\\
                       0 & 0 & \frac{l^2}{2}+\mathbb{E}\\
                       0 & \frac{l^2}{2}-\mathbb{E} & 0
                      \end{array} \right),~~~~F^{^{(1)}}_{a j}~=~0,~~~~F^{^{(2)}}_{i b}~=~0,~~~~F_{ij} =l^2,
\end{eqnarray}
we obtain the background of the original $\sigma$-model. It is given by
\begin{eqnarray}
ds^2&=&l^2\big(\frac{1}{4} dx_{_{1}}^2 + e^{-x_{_{1}}} d x_{_{2}} d x_{_{3}} + d z^2\big),\label{VI.36}\\
{ B} &=& \mathbb{E} e^{-x_{_{1}}} \big(-x_{_{2}} d x_{_{1}} \wedge d x_{_{3}} + d x_{_{2}} \wedge d x_{_{3}}\big),\label{VI.37}
\end{eqnarray}
where $z$  stands for the coordinate of the orbit $O$ of $G$.
The metric \eqref{VI.36} can be written as $ds^2 =ds^2_{_{AdS_{_3}}} + l^2 dz^2$.
In order to get more insight of this metric one may use the following transformation
\begin{eqnarray}
e^{\frac{x_{_{1}}}{2}}=\frac{1}{\rho},~~~~~~  ~~~~~ x_{_{2}}=\frac{1}{l}(\tau-x),~~~~~~~~~ x_{_{3}} =-{l}(\tau+x),\label{VI.38}
\end{eqnarray}
then, the metric becomes the same as the $AdS_3 \times \mathbb{R}$ space given by \eqref{b.11.2}.
As explained in Sec. \ref{Sec.III}, the metric \eqref{b.11.2} is locally equivalent to the case of
$\beta =1$ of the G\"{o}del metric \eqref{b.10}.
In addition, one easily gets that the field strength corresponding to the $B$-field \eqref{VI.37} is zero.
We have checked that the metric constructed on the double $(sl(2, \mathbb{R}) , 3{\cal A}_1)$ with a vanishing field strength
can't be conformally invariant up to the one-loop order.
Therefore, according to Ref. \cite{N.Mohammedi} we don't expect to have a conformally invariant dual theory at one-loop level.
One immediately verifies the field equations \eqref{a.2}-\eqref{a.4}
for the metric \eqref{VI.36} \footnote{Considering $z$ as the time coordinate in \eqref{VI.36} one obtains an anisotropic homogeneous spacetime
based on the $SL(2 , \mathbb{R})$ Lie group so that it is a solution for two-loop beta-function equations. Recently,
the two-loop beta-function equations with dilaton and $B$-field on the anisotropic homogeneous
spacetimes of the Bianchi-type have been investigated \cite{F.Naderi}.} and vanishing field strength together with the dilaton
field ${ \Phi}=f z+b$ for which $f^2 = 2 l^2 \Lambda -3$;
moreover, to satisfy the field equations we must have a coupling constant in the form of
$\alpha' = l^2$.

In the same way, to construct out the dual $\sigma$-model on the manifold
$\tilde {{\cal M}} \approx O \times \tilde {G}$ with the Lie group  $3A_1$
we parameterize the corresponding Lie
group  with  coordinates $({\tilde x}_{_{1}} , {\tilde x}_{_{2}} , {\tilde x}_{_{3}})$ so that its element
is defined as \eqref{VI.33} by replacing untilded quantities with tilded ones.  Utilizing the relation \eqref{VI.32} and also
\eqref{c.4} for tilded quantities we get
\begin{eqnarray}\label{VI.40}
{\tilde \Pi}_{_{ab}}(\tilde g)\;=\;\left( \begin{array}{ccc}
                     0 & -{\tilde x}_{_2}&{\tilde x}_{_3}\\

                     {\tilde x}_{_2} & 0 &-2 {\tilde x}_{_1}\\

                     -{\tilde x}_{_3} &  2{\tilde x}_{_1}& 0
                      \end{array} \right).
\end{eqnarray}
Finally, by using \eqref{c.11} the metric and $\tilde B$-field of the dual model take the following forms
\begin{eqnarray}
d\tilde s^2&=&l^2 d z^2+\frac{4}{{\tilde \Delta}}\Big[l^2(d{\tilde x_{_1}}'^2 +d {\tilde x_{_2}} d {\tilde x_{_3}})
-\frac{4}{l^2} \Big(2{{\tilde x_{_1}}}' d {{\tilde x_{_1}}}'+ d \big({\tilde x_{_2}}  {\tilde x_{_3}}\big)\Big)^2\Big],\label{VI.41}\\
{ \tilde B} &=& \frac{8}{{\tilde \Delta}}\Big[{\tilde x_{_3}} d {\tilde x_{_1}}' \wedge  d {\tilde x_{_2}}
-{\tilde x_{_2}} d {\tilde x_{_1}}' \wedge  d {\tilde x_{_3}}
+ {\tilde x_{_1}}' d {\tilde x_{_2}} \wedge  d {\tilde x_{_3}}\Big],~~~~\label{VI.42}
\end{eqnarray}

\begin{table}
	\caption{ Non-Abelian T-duality of the G\"{o}del string cosmologies without corrections}
	\centering
	\begin{tabular}{| p{4em} | p{21em} | p{6.75em} | p{7.25em}|}
		\hline
		\multicolumn{2}{|c|}{\scriptsize Non-Abelian T-duality  with the $A_2$ Lie group}  & \scriptsize  Conditions for the one-loop solution & \scriptsize Conditions for the two-loop solution\\
		\hline

		\multirow{3}{4em}{\scriptsize Original background}&{\scriptsize $ds^2=l^2\big[-d{y_{_{1}}}^2+d{y_{_{2}}}^2 + d{x_{_{1}}}^2+(\beta-1)e^{2x_{_{1}}}d{x_{_{2}}}^2-2e^{x_{_{1}}}d{y_{_{1}}}d{x_{_{2}}}\big]$}&  \multirow{3}{6.75em}{-----}   & \multirow{3}{7.25em}{\scriptsize $\beta\neq1$ with the conditions given in Eqs. \eqref{b.18} and \eqref{b.19} } \\
		
		&{\scriptsize $B=2e^{x_{_{1}}}\big(\mathbb{E}d{x_{_{2}}}\wedge d{y_{_{1}}}+\mathbb{N}d{x_{_{2}}}\wedge d{y_{_{2}}}\big)$} && \\
		
		&{\scriptsize $ \Phi=b,~~~~\beta\neq 1$}&& \\
		
		\hline

		\multirow{6}{4em}{\scriptsize Dual background}&\multirow{3}{21em}{\scriptsize ${d\tilde {s}}^2=\frac{1}{\tilde\Delta}\Big[-l^2({\tilde x_{_2}}^2 +\beta l^4 -4\mathbb{E}^2) d{t}^2 +
			l^2\big({\tilde x_{_2}}^2 +(\beta-1) l^4 +4\mathbb{N}^2\big) d{z}^2 +l^2(\beta-1) d {\tilde x_{_1}}^2+l^2 d {\tilde x_{_2}}^2 -2 l^2  {\tilde x_{_2}}
			d {\tilde x_{_1}} dt + 4\mathbb{E}l^2 d {\tilde x_{_2}} dt+ 4\mathbb{N}l^2 d {\tilde x_{_2}} dz +8\mathbb{E}\mathbb{N}l^2 d z dt\Big]$}&\multirow{6}{6.75em}{-----}   &\multirow{6}{6.75em}{-----} \\
		
		&&& \\
		
		&  & & \\
		
		&\scriptsize ${\tilde B} = \frac{1}{\tilde\Delta}\Big[{\tilde x_{_2}} d {\tilde x_{_1}} \wedge  d {\tilde x_{_2}} + 2\mathbb{E} {\tilde x_{_2}} d {\tilde x_{_1}} \wedge  dt
		+ 2\mathbb{N} {\tilde x_{_2}} d {\tilde x_{_1}} \wedge  dz  -l^4  d {\tilde x_{_2}} \wedge  dt
		- 2\mathbb{N} l^4  d z \wedge  dt\Big]$& & \\
		&\scriptsize $\tilde\Delta=\tilde {x}_{_2}^2+(\beta-1)l^4$& &\\
		\hline
	\end{tabular}
	
	\begin{tabular}{| p{4em} | p{21em} | p{6.75em} | p{7.25em}|}
		\hline
		\multicolumn{2}{|c|}{\scriptsize Non-Abelian T-duality  with the $A_2\oplus A_1$ Lie group}  & \scriptsize  Conditions for the one-loop solution & \scriptsize Conditions for the two-loop solution\\
		\hline

		\multirow{3}{4em}{\scriptsize Original background}&{\scriptsize $ds^2=l^2\big[-dt^2+d{x_{_{1}}}^2 + d{x_{_{3}}}^2+(\beta-1)e^{2x_{_{1}}}d{x_{_{2}}}^2-2e^{x_{_{1}}}dtd{x_{_{2}}}\big],$}&  \scriptsize $\beta=1,$   &\scriptsize $\mathbb{E}^2=\frac{l^4(4\beta-3)}{4},$  \\
		
		&{\scriptsize $B=2e^{x_{_{1}}}\big(\mathbb{E}d{x_{_{2}}}\wedge dt+\mathbb{N}d{x_{_{2}}}\wedge d{x_{_{3}}}+\frac{s}{2}d{x_{_{1}}}\wedge d{x_{_{2}}}\big),$} &\scriptsize $\mathbb{N}=0,~\mathbb{E}^2=\frac{l^4}{4},$&\scriptsize $\alpha'=\frac{2l^2\beta}{5-6\beta},~\mathbb{N}=0$,\\
		
		&{\scriptsize $ \Phi=fx_{_{3}}+b$}&\scriptsize $f^2=2l^2\Lambda-1$&\scriptsize $f^2=2(1+l^2\Lambda-\frac{1}{\beta})$ \\

		\hline

		\multirow{7}{4em}{\scriptsize Dual background}&\multirow{3}{21em}{\scriptsize ${d\tilde {s}}^2=\frac{1}{\hat{\Delta}}\Big\{-l^2\big[(s-{\tilde x_{_2}})^2 +\beta l^4 +4(\mathbb{N}^2-\mathbb{E}^2)\big] d{t}^2~~~~+\frac{1}{l^2} \big[4\mathbb{N}^2 +l^4(\beta-1)\big] d {\tilde x_{_1}}^2+l^2 d {\tilde x_{_2}}^2+\frac{1}{l^2} \big[(s-{\tilde x_{_2}})^2+ l^4(\beta-1)\big]d {\tilde x_{_3}}^2 -4\mathbb{N} l^2  d {\tilde x_{_3}} dt + 4\mathbb{E}l^2 d {\tilde x_{_2}} dt+2 l^2  (s-{\tilde x_{_2}}) d {\tilde x_{_1}} dt+\frac{4}{l^2} \mathbb{N} (s-{\tilde x_{_2}}) d {\tilde x_{_1}} d {\tilde x_{_3}} \Big\},$}&   &  \\
		
		&&\scriptsize $\beta=1,~\mathbb{E}^2=\frac{l^4}{4}$  &\scriptsize $\beta=1,~\mathbb{E}^2=\frac{l^4}{4}$ \\
		&  &\scriptsize $\tilde f^2=\frac{2l^2\tilde\Lambda-1}{l^4},$ &\scriptsize $\tilde f^2=\frac{2l^2\tilde\Lambda-1}{l^4},~\mathbb{N}=0,$\\
		
		&\multirow{2}{21em}{\scriptsize ${\tilde B} =\frac{1}{\hat{\Delta}}\Big\{({\tilde x_{_2}}-s) d {\tilde x_{_1}} \wedge  d {\tilde x_{_2}} + 2\mathbb{E} ({\tilde x_{_2}}-s) d {\tilde x_{_1}} \wedge  dt
			- 2\mathbb{N} d {\tilde x_{_2}} \wedge  d {\tilde x_{_3}}-l^4  d {\tilde x_{_2}} \wedge  dt+4\mathbb{E}\mathbb{N}  d {\tilde x_{_3}} \wedge  dt\Big\},$}&\scriptsize $\mathbb{N}=0,~\tilde {x}_{_2}\rightarrow \tilde {x}_{_2}+s$ &\scriptsize $\tilde {x}_{_2}\rightarrow \tilde {x}_{_2}+s$ \\
		&&& \\
		&\scriptsize $\tilde{\Delta}=(s-{\tilde x_{_2}})^2 +4\mathbb{N}^2+(\beta-1)l^4$& &\\
		&\scriptsize $\tilde\Phi={\tilde f} \tilde {x}_{_3}+b$&& \\
		\hline
	\end{tabular}
	
	\begin{tabular}{| p{4em} | p{21em} | p{6.75em} | p{7.25em}|}
		\hline
		\multicolumn{2}{|c|}{\scriptsize  Non-Abelian T-duality  with the $SL(2,\mathbb{R})$ Lie group}  & \scriptsize  Conditions for the one-loop solution & \scriptsize Conditions for the two-loop solution\\
		\hline

		\multirow{3}{4em}{\scriptsize Original background}&{\scriptsize $ds^2=l^2\big[\frac{1}{4}d{x_{_{1}}}^2 +e^{-x_{_{1}}} d{x_{_{2}}}d{x_{_{3}}}+dz^2\big]$}& \multirow{3}{6.75em}{-----}  &\scriptsize $\alpha'=l^2,$  \\
		
		&{\scriptsize $B=\mathbb{E} e^{-x_{_{1}}} \big(-x_{_{2}} d x_{_{1}} \wedge d x_{_{3}} + d x_{_{2}} \wedge d x_{_{3}}\big)$} &&\scriptsize $f^2=2l^2\Lambda-3$\\
		
		&{\scriptsize $ \Phi=fz+b$}&& \\

		\hline

		\multirow{4}{4em}{\scriptsize Dual background}&{\scriptsize ${d\tilde {s}}^2=l^2 d z^2+\frac{4}{{\tilde \Delta}}\Big[l^2(d{\tilde x_{_1}}'^2 +d {\tilde x_{_2}} d {\tilde x_{_3}})-\frac{4}{l^2} \Big(2{{\tilde x_{_1}}}' d {{\tilde x_{_1}}}'+d \big({\tilde x_{_2}}  {\tilde x_{_3}}\big)\Big)^2\Big],$}&\multirow{4}{6.75em}{-----}  & \multirow{4}{7.25em}{-----} \\
		
		&{\scriptsize ${\tilde B} =\frac{8}{{\tilde \Delta}}\Big[{\tilde x_{_3}} d {\tilde x_{_1}}' \wedge  d {\tilde x_{_2}}-{\tilde x_{_2}} d {\tilde x_{_1}}' \wedge  d {\tilde x_{_3}}+ {\tilde x_{_1}}' d {\tilde x_{_2}} \wedge  d {\tilde x_{_3}}\Big],$}& & \\
		&\scriptsize ${\tilde \Delta}=l^4 -16({{\tilde x_{_1}}}'^2 + {\tilde x_{_2}} {\tilde x_{_3}})$ & &\\
		
		&\scriptsize $\tilde {x}'_{_1}=\frac{\mathbb{E}}{2}-\tilde {x}_{_1}$&& \\
		
		\hline
	\end{tabular}
\end{table}
where ${\tilde \Delta}=l^4 -16({{\tilde x_{_1}}}'^2 + {\tilde x_{_2}} {\tilde x_{_3}})$ and ${\tilde x_{_1}}' = \mathbb{E}/2-{\tilde x_{_1}}$.
For the ${\tilde B}$-field there is only a nonzero component of the field strength, obtaining
\begin{eqnarray}
{\tilde H}_{_{{\tilde x_{_1}}' {\tilde x_{_2}} {\tilde x_{_3}} }} = \frac{4}{{\tilde \Delta}^{^2}}
\big[3l^4 -16 ({\tilde x_{_1}}'^2+ {\tilde x_{_2}} {\tilde x_{_3}})\big]. \label{VI.43}
\end{eqnarray}
As mentioned above, since the background of the original $\sigma$-model doesn't satisfy the one-loop beta-function equations,
the dual background can't be conformally invariant at the one-loop order.
We have represented the results of the non-Abelian T-duality of the G\"{o}del string cosmologies in Table 3.


\section{Conclusion}

We have obtained some new solutions for the field equations of
bosonic string effective action up to  first order in $\alpha'$, including
the G\"{o}del spacetimes, axion field and dilaton.
Our results have shown that these solutions can be appropriate to study (non-)Abelian T-dualization of G\"{o}del
string cosmologies via PL T-duality approach.
In studying Abelian duality of the G\"{o}del spacetimes we have found seven dual models
in a way that the models are constructed by one-, two- and three-dimensional Abelian Lie groups acting freely on the target space manifold.
When the dualizing was implemented by the shift of directions $t$, $z$ and $(t, z)$ we showed that the pair of the mutually dual models can be
conformally invariant at one-loop level, in a way that the corresponding dual dilaton field was found by using transformation
\eqref{c.16.3}.
Our results showed that the Abelian T-dual models are,
under some of the special conditions, self-dual; moreover, by applying the rules of Abelian
T-duality without further corrections, we were still able to obtain two-loop solutions.
We also studied the Abelian T-duality of G\"{o}del string cosmologies up to $\alpha'$-corrections
by using the T-duality rules at two-loop order of KM.
Most importantly, we have obtained the non-Abelian duals of the G\"{o}del
spacetimes. First, we have constructed the T-dual models on
the four-dimensional manifold ${ {\cal M}} \approx O \times  {G}$ with two-dimensional non-Abelian Lie group and have shown that
the metric of the dual model has true singularities for the range of $0 < \beta < 1$. In this case, the models
are valid for all values of $\beta$  except for $1$.
Unfortunately, the dual model doesn't satisfy the two-loop beta-function equations.
We have then found other non-Abelian duals for the G\"{o}del spacetimes by applying
the $A_2 \oplus A_1$ and $SL(2, \mathbb{R})$ Lie groups. The case of $\beta=1$ of dual model constructed on
the semi-Abelian double $({\cal A}_2 \oplus {\cal A}_1 , 3{\cal A}_1)$  is
conformally invariant to zeroth order in $\alpha'$, as well as to the first order.
In this way, it has shown that the $\beta=1$ case of the pair of the mutually dual models
as the $AdS_3 \times \mathbb{R}$ spaces are PL self-dual.
Finally, we have shown that the model constructed by the double $(sl(2, \mathbb{R}) , 3{\cal A}_1)$
leads to the case of $\beta=1$ of the  G\"{o}del metrics with zero field strength. Indeed,
the model didn't satisfy the one-loop beta-function equations.
Because of the vanishing traces of the structure constants corresponding to the double
$(sl(2, \mathbb{R}) , 3{\cal A}_1)$, the dual model couldn't be also conformally invariant at the one-loop order.


\subsection*{Acknowledgements}

This work has been supported by the research vice
chancellor of Azarbaijan Shahid Madani University under research fund No. 97/231.
A. Eghbali is especially grateful to A. Mehrvand for his careful reading of the manuscript.


\end{document}